%% file: pulsar.tex
\documentclass[fleqn,usenatbib]{mnras}

\usepackage{newtxtext,newtxmath}

\usepackage[T1]{fontenc}

\DeclareRobustCommand{\VAN}[3]{#2}
\let\VANthebibliography\thebibliography
\def\thebibliography{\DeclareRobustCommand{\VAN}[3]{##3}\VANthebibliography}

\usepackage{graphicx}	
\usepackage{amsmath}	
\usepackage{rotating}
\usepackage{hyperref}
\usepackage{gensymb}
\usepackage{url}
\usepackage{booktabs}
\usepackage{longtable}

\newcommand{\rxte}{\textit{RXTE}}
\newcommand{\pca}{\textit{RXTE}/PCA} 

\newcommand{\obj}{V~0332+53}

\newcommand{\ironka}{iron~$K_{\alpha}$}
\newcommand{\ergps}{erg\,s$^{-1}$}
\newcommand{\ergpspcm}{erg \,s$^{-1}$\,cm$^{-2}$}

\title[Iron spectral features in \obj]{Pulsating iron  spectral features in the emission of  X-ray pulsar \obj}

\author[Bykov et al.]{
S.D. Bykov$^{2,1}$\thanks{E-mail:sergei.d.bykov@gmail.com}, E.V. Filippova$^{1}$, M.R. Gilfanov$^{2,1}$, S.S. Tsygankov$^{4,1}$, A.A. Lutovinov$^{1,3}$ \newauthor and S.V. Molkov$^{1}$
\\
$^{1}$Space Research Institute, Russian Academy of Sciences, Profsoyuznaya 84/32, 117997 Moscow, Russia\\
$^{2}$Max Planck Institute for Astrophysics, Karl-Schwarzschild-Str 1, Garching b. München D-85741, Germany\\
$^{3}$Higher School of Economics, Myasnitskaya 20, 101000 Moscow, Russia \\
$^{4}$Department of Physics and Astronomy, FI-20014 University of Turku,  Finland}

\date{Accepted XXX. Received YYY; in original form ZZZ}

\pubyear{2020}

\begin{document}
\label{firstpage}
\pagerange{\pageref{firstpage}--\pageref{lastpage}}
\maketitle

\begin{abstract}
We present results of phase- and time-resolved study of iron spectral features in the emission of the Be/X-ray transient pulsar \obj\,  during its type II outburst in 2004 using archival \pca\,  data. Coherent pulsations  of both  fluorescent iron line at $\approx 6.4$ keV and neutral iron K-edge at $\approx 7.1$ keV have been detected throughout the entire outburst. The pulsating iron K-edge is reported for the first time for this object. Near the peak of the outburst, the 3--12 keV pulse profile shows two deep, $F_{\rm max}/F_{\rm min}\sim 2$, and narrow dips of nearly identical shape, separated by exactly $\Delta\phi=0.5$ in phase. The dip spectra are nearly identical to each other and very similar in shape to the spectra outside the dips.  The iron K-edge peaks at the phase intervals corresponding to the dips, although its optical depth $\tau_K\sim 0.05$ is by far insufficient to explain the dips. The iron line shows pulsations with a complex pulse profile without any obvious correlation with the total flux or optical depth of the K-edge.  Accounting for the component associated with  reprocessing of the pulsar emission by the surface of the donor star and circumstellar material, we find a very high  pulsation amplitude of the iron line flux, $F_{\rm max}/F_{\rm min}\sim 10$. We demonstrate that  these properties of \obj\ can not be explained by contemporary emission models for accreting X-ray pulsars and speculate about the origin of the observed iron spectral features.

\end{abstract}

\begin{keywords}
pulsars: individual: V 0332+53 –X-rays: binaries.
\end{keywords}



\section{Introduction}\label{par:intro}
Be/X-ray binaries (BeXRBs) are binary systems  harbouring a neutron star and a fast-spinning early-type star with an equatorial circumstellar disc \citep{Reig2011}. Such objects are known for their transient behaviour in X-rays and show two types of outbursting activity \citep{Reig2011}. Type I outbursts  happen periodically when the neutron star passes a periastron of its eccentric orbit \citep{Okazaki2002}, with the maximum luminosity reaching up to $10^{37} $\ergps. On the other hand, giant (Type II) events are rare, not related to any orbital phase and much brighter than type I (achieving or even surpassing the Eddington luminosity limit for a neutron star $\sim10^{38}$\,~\ergps). The origin of giant outbursts is not exactly known  and is probably related to mass ejection events from the companion star \citep{Okazaki2001}.  Due to the large range of observed luminosities during type II outbursts, these objects permit to explore the structure of the accretion flow around magnetised neutron star in a broad range of the mass accretion rates. Examining the matter distribution close to the neutron star allows studying  the  interaction of matter with high magnetic and radiation fields near the pulsar and the shape of the accretion structures formed as a result of this interaction. Besides, the accretion regimes may change with luminosity, which, in turn, lead to the changes in beaming patterns of the X-ray emission from the accreting neutron star \citep[eg][]{Basko1976}. Such changes directly affect what an observer sees, hence serving as a probe of the complex physics of magnetized accretion.

In binary systems, some fraction   of the primary X-ray emission may be intercepted by the surrounding relatively cold matter  which leads to the appearance of the  reprocessed emission.
This emission, in particular  fluorescent iron $K_\alpha$ (2P-1S) line, 
 can serve as a powerful tool to study  the spatial distribution and ionization state of the material around the X-ray sources \citep{Basko1974, Inoue1985, Makishima1986, Fabian1989, George1991, Gilfanov1999, Gilfanov2010, Tsygankov2010b, Gimenez-Garcia2015, Aftab2019}.
 The energy of this  line is $\approx 6.4$ keV for neutral and low-ionized iron atoms and increases up to $\approx 7$ keV for H-like iron, whilst the corresponding K-absorption edge  energy varies between $\sim 7.1$ and $\sim9$ keV respectively.
Iron has a large fluorescent yield ($\sim$30\% for low-ionized ions) and its $K_\alpha$ line energy  falls in the standard X-ray band, i.e. within the sensitivity range of a plethora of X-ray instruments.  The flux,  equivalent width (EW) and  shape of the iron line depend on the relative location of the X-ray source and the reflecting medium, its area, density, kinematics  and ionization state. For example, significant progress has been made by applying the 'iron line tomography' to the High-Mass X-ray binaries (HMXB) \citep[e.g.][]{Nagase1992,Day1993, Gimenez-Garcia2015, Aftab2019}  as well as low mass X-ray binaries \citep{Gilfanov1999, Gilfanov2000, Churazov2001}.  In HMXB X-ray pulsars,  the list of candidates   for the fluorescent emission production sites  includes all main components of the accretion flow: the accretion column, the accretion stream, the Alfven surface (shell),  accretion disc, the stellar wind material, as well as the surface of the massive donor star itself \citep[e.g., ][]{Inoue1985}.

In some HMXB pulsars, the equivalent width of the iron line  was shown  to  vary with the rotational phase of the neutron star.  Among others, the pulsating iron line was detected in  LMC X-4 \citep{Shtykovsky2017}, Cen X-3 \citep{Day1993},  GX 301-2 \citep{Liu2018,Zheng2020}, Her X-1 \citep{Choi1994} and   4U 1538-522 \citep{Hemphill2014}. Interestingly, none of these sources is a BeXRB.

Several models were proposed to explain these variations. It was suggested that  variations in the line equivalent width may be caused by the variations in time of the observed columns density of the cold material  in the vicinity of the neutron star \citep[e.g.][and references therein]{Inoue1985, Leahy1989}.  On the other hand, the pulsating nature of the iron line may indicate that the matter is not distributed symmetrically around the pulsar. In this case, the irradiating flux from the primary varies with the rotation of the pulsar  \citep{Inoue1985, Day1993}.  \citet{Shtykovsky2017} used  phase-resolved spectroscopy to study LMC X-4 and concluded  that the iron line emission possibly comes from the hotspot on the accretion disc.  Studying changes in the emission pattern during the eclipses of the HMXB pulsar Cen X-3 \citet{Nagase1992} found that the 6.4 keV line must be produced fairly close to the neutron star and obtained  an upper limit for the distance between the fluorescent matter and the X-ray source of $\sim  1$ lt-sec (light second).
  Later,  \citet{Kohmura2001} proposed for this source that the reprocessing site is the accreting matter flowing along the magnetic field lines at  the distance of $\approx 1.7 \times 10^8$ cm from the neutron star. \citet{Sanjurjo-Ferin2021} propose the accretion stream in Cen X-3 as a source of reprocessed neutral iron. Similarly, in GX 301-2  \citet{Liu2018} detected transient pulsations of iron line flux, suggesting that the reprocessing medium might be the accretion stream or the secondary's surface. On the other hand, \citet{Zheng2020} argued that the size of the emission region  in this pulsar varies in the range of $\sim 0\textrm{--}40$ light seconds, with the average values consistent with the distance from the pulsar to the accretion stream from the secondary star.  
  \citet{Endo2002} found that some fraction of the iron line emission in GX 301-2 originates in the  accretion column itself.  Recently \citet{Yoshida2017} proposed that the modulation of the iron line intensity may be due to the finite speed of light, and the effect is determined by the size of reprocessing region. 
  
 An important clue for the origin of the iron line may be provided by its evolution during outbursts of transient pulsars. However, evidence is still incomplete and controversial.  In 4U 0115+63, \citet{Tsygankov2007} found that the equivalent width of the iron line decreases  with the declining luminosity of the pulsar, which was interpreted as the decrease of the solid angle of the reprocessing material as seen by the primary emission source.   
  For 1A 1118-615, on the contrary, it was shown that the equivalent width of the line was constant during the type II outburst \citep{Nespoli2011}.

   A promising and so far unexplored venue is the  spectroscopy of the iron absorption K-edge. Recently, \citet{Yoshida2019} proposed the accretion stream from the inner disc onto pulsar (the accretion curtain) as a source of low-ionized fluorescent emission based on the dynamics of iron K-edge absorption in Vela X-1, GX 1+4 and, possibly, in two other pulsars. Their arguments were based on the variations of the K-edge depth as well as the presence of  dips in the observed count rate (flux) of the sources related to the eclipse of the emitting region by the accretion column. They suggested that the matter captured by the magnetic field  which co-rotates with the neutron star can produce the observed  variability in K-edge absorption optical depth.  To our knowledge, this is the first and so far the only detection of the variability of the iron absorption edge in accretion-powered X-ray pulsars.
  
  Thus, a wealth of information is provided by the X-ray spectral features associated with iron absorption and fluorescence. This information may help to constrain the geometry of the accretion column, accretion flow and surrounding material and shed further light on the emission mechanisms in accreting X-ray pulsars. In the present paper, we analyse the behaviour of iron spectral features in  Be/X-ray HMXB pulsar \obj\, during its type II outburst in 2004.

\subsection*{System \obj}

The luminous transient X-ray pulsar \obj\, was discovered in 1973 during its bright outburst  with a peak  intensity of $\sim1.4$ Crab in $3\textrm{--}12$ keV energy band \citep{Terrell1984}.
The pulsation period  of $\sim$4.4 s and parameters of orbital motion (orbital period of $\sim$ 34 days and eccentricity of $\sim$0.3) were determined later, during the next less prominent outbursts in 1983-1984 \citep{Stella1985}.
The optical counterpart was found to be an O8-9Ve star BQ Cam \citep{Honeycutt1985}. 
Another outburst of the source was registered in 1989 \citep{Makishima1990}. The significant cyclotron resonance scattering feature (CRSF) at the energy 28.5 keV was detected, which allowed estimating the magnetic field strength to be around $\sim 2.5 \times 10^{12}$ G.

Based on the optical properties of the source the distance to the system was initially estimated to be $\sim7$ kpc \citep{Negueruela1999}. More recently,  \citet{RoucoEscorial2019}  and \citet{Arnason2021} used  Gaia DR2 measurements and derived the distance of  $5.1^{+0.9}_{-0.7}$ kpc and $5.1^{+0.8}_{-1.0}$ respectively. We thus note that majority of initial studies of \obj\ adopted the distance of 7 kpc and therefore  overestimated the source luminosity by a factor of $\sim$2,  if the   true distance to the pulsar is $\approx 5$ kpc.  In this work we adopt the Gaia distance of 5.1 kpc.

The next major outburst of \obj\, occurred in 2004 and was fully covered by \rxte\, observations. Two CRSF harmonics we detected in the source spectrum, which showed  complex evolution in time and with luminosity \citep[see e.g.,][]{Kreykenbohm2005,Tsygankov2006,Tsygankov2010,Lutovinov2015}. 
The observed anti-correlation between the  CRSF centroid energy and  luminosity during the outburst can be understood  either as the changing height of the accretion column  \citep{Tsygankov2006,Mushtukov2015}, or as a result of reflection of the accretion column emission from the surface of the neutron star \citep{Poutanen2013, Lutovinov2015,Mushtukov2018}. 
The behaviour of the iron line emission was studied succinctly in \citet{Tsygankov2010b}. They reported the modulation of iron line flux and equivalent width with the rotation of the pulsar. They also found long term variability of the equivalent width of the iron line which seemed to correlate with the orbital phase and tentatively interpreted this fact as fluorescence of the material at the surface of the optical companion or its circumstellar disc.

The most recent giant outburst of the source took place in 2015, followed by a mini-outburst in 2016 \citep{Cusumano2016, Tsygankov2016,Baum2017}. In \citet{Doroshenko2017, Vybornov2018} it was found that the above-mentioned  anti-correlation between  CRSF energy and  luminosity breaks at low fluxes, suggesting the change in accretion regimes \citep[from super- to sub-critical, see][]{Basko1976, Mushtukov2015b}.
Based on all observational data up to 2015, \citet{Doroshenko2016} obtained the orbital solution of the pulsar which we use in this work (Table \ref{tab:orb_pars}). The evolution of the rotational frequency  of \obj\ with luminosity proposes that the pulsar is accreting from the disc \citep{Doroshenko2016,Filippova2017} and not the wind.

The goal of the present  paper is to utilize the information provided by the spectral features of iron to study the structure of the accretion flow in the vicinity of the neutron star as well as the overall distribution of the circumstellar material  in the system \obj. The paper is based on  \rxte\ observations of the  type II outburst of the source in 2004. The paper is structured as follows. 
The details of \rxte\,  data reduction are presented in sect. \ref{par:data_analysis}. In sect. \ref{par:phase_ave} we investigate the evolution of the flux and equivalent width of the iron line and of the depth of its K-edge on the time-scales of the outburst. The pulse-phase-resolved modulations of the iron line and K-edge parameters are studied in sect. \ref{par:phase_resolved}. 
In sect. \ref{par:discussion} we discuss our results and constrain the location of the material responsible for absorption and emission features in the source spectrum.  Our conclusions are summarized in sect. \ref{par:conclusion}

\section{Observations and data analysis}\label{par:data_analysis}
During its 2004-2005 type II outburst, \obj\, was monitored by \textit{Rossi X-ray Timing Explorer} (\rxte)\, observatory, covering all phases of the outburst. In this work we use all available \pca\, (Proportional Counter Array) spectrometer data between MJD 53336 and MJD 53440,  97 observations in total from proposals 90014, 90089, 90427. 

\pca\, spectrometer \citep{Jahoda1996} is an array of five proportional counters sensitive in the energy range 3-60 keV. It has a total collecting area of 6400 cm$^2$ (for all 5 units) and $20\%$ energy resolution at 6 keV (depending on the energy binning of configurations). 
We reduced  \pca\, data following the \rxte\, cookbook\footnote{\url{https://heasarc.gsfc.nasa.gov/docs/xte/recipes/cook_book.html}} with the standard FTOOLS/HEASOFT v 6.24 package.  Observation-averaged spectra were extracted from the top layer of PCU2 detector (the best calibrated one) in Standard2 data-mode,  an additional systematic error of $0.25\%$ was added on all channels due to uncertainties in the telescope's response \citep{Garcia2014}. All spectra were fit with {\sc xspec}  package \citep{Arnaud1996} using $\chi^2$ statistics. 

\begin{table}
    \caption{Orbital parameters of \obj\, \citep{Doroshenko2016}}
    \label{tab:orb_pars}
    \centering
\begin{tabular}{l|c}
Parameter   & Value        \\ \hline
$P_{\rm orb}$ - orbital period   & 33.850(1) days   \\
$a\,sin(i)$ - semi major axis projection on the line of sight & 77.81(7) lt s   \\
e - eccentricity          & 0.3713(8)       \\
$\omega$ -longitude of periastron   & 277.43(5) deg   \\
$T_{\rm PA}$ - periastron time passage   & MJD 57157.88(3)
\end{tabular}

\end{table}

There were 72 observations that had data configuration appropriate for phase-resolved spectroscopy in energy range $3\textrm{--}12$ keV  (data modes E\_125us\_64M\_0\_1s,  B\_16ms\_46M\_0\_49\_H and B\_16ms\_64M\_0\_249). The time of arrival of photons was corrected to solar system barycentre,  and the Doppler shift due to orbital motion of a pulsar in the binary was corrected for using  orbital parameters from \citet{Doroshenko2016} (see Table \ref{tab:orb_pars}). In the phase-resolved analysis, photons were folded into 16, 12 or 8 phase bins using  \textit{fasebin} FTOOLS/HEASOFT task. In the last four observations (90014-01-08-(00-03)), pulsations were not detected, correspondingly no phase-resolved analysis was performed.   
 
  All but three observations during the rising part of the outburst (up to MJD 53360) had E\_125us\_64M\_0\_1s configuration (the remaining  three  were in B\_16ms\_64M\_0\_249 data mode). Between MJD 56383 and 53403 (the declining part of the outburst) the data suitable for phase-resolved spectroscopy had B\_16ms\_46M\_0\_49\_H configuration, and after MJD 53403 it had E\_125us\_64M\_0\_1s data mode again. 
 Unfortunately, the  E\_125us\_64M\_0\_1s configuration in PCU2 has a corrupt channel 10 covering $\sim6.14\textrm{--}6.54$ keV energy range, which registers no photons. As these energies are of primary interest for our study we had to exclude  observations made in  this configuration from phase-resolved spectroscopy of the iron line.   In phase-resolved spectra, a systematic error of $0.4\%$ was added to all channels. The systematic error  differs from the one used in Standard2 mode because  the data from all three layers of active PCUs were gathered into one event file in these configurations (B\_16ms\_46M\_0\_49\_H and B\_16ms\_64M\_0\_249).  
 
 Observations with similar pulse profiles and the same configuration/detectors  were combined  using \textit{fbadd} task, and their phase-resolved spectra were combined  with \textit{addspec} script (see sect. \ref{par:phase_resolved} for details about groups).
 
 All that leaves us with 30 observations and 8 groups where we performed phase-resolved spectroscopy.

\begin{figure}
    \centering
    \includegraphics[width=0.5\textwidth]{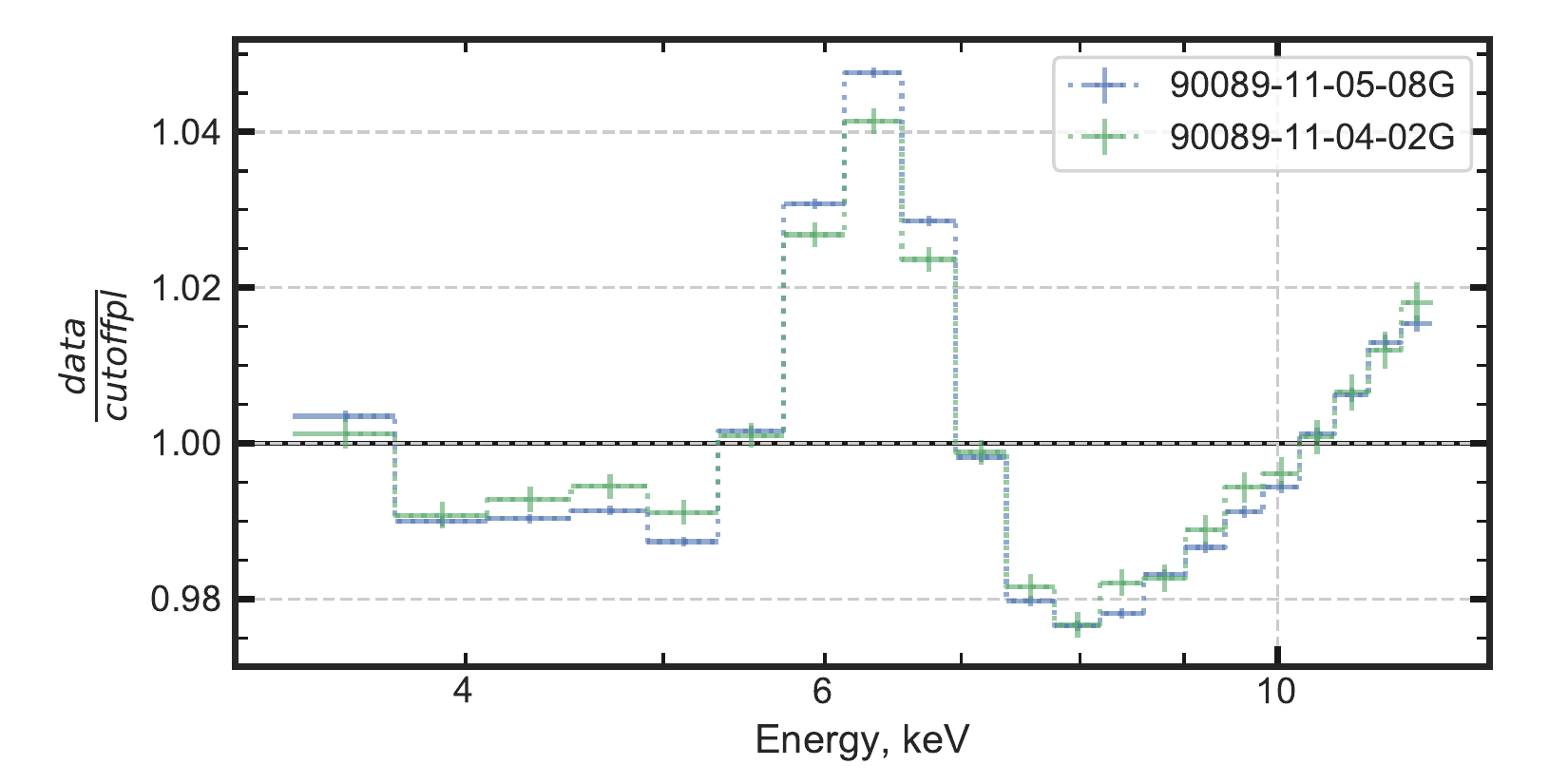}
    \caption{The ratio of data to  exponentially cutoff power-law  in the time-averaged spectra of two observations (90089-11-05-08G, MJD 53365.3 and 90427-01-03-00, MJD 53384.4).}
    \label{fig:rxte-ratio}
\end{figure}

\section{Results}
\subsection{Light curve and long-term spectral evolution}
\label{par:phase_ave}

\citet{Tsygankov2010b} reported  the presence of a strong iron emission line ($EW\sim60-100$ eV) in the spectra of \obj. To corroborate this result we plot the ratio of  the source spectra obtained in two observations around MJD 53365.3 and 53384.4 to the best-fitting model of a power-law with exponential cutoff (Fig.~\ref{fig:rxte-ratio}). The plot reveals  a strong emission feature at $\sim6.4$ keV as well as a  depression in the spectrum near $\sim7$ keV which resembles the behaviour observed in  black hole X-ray binaries \citep[e.g.][]{Gilfanov1999}. 
Similar to the latter, these features can be associated with the fluorescent K-$\alpha$ line and absorption K-edge of low-ionized iron and, possibly, the lower energy part of the Compton reflection bump.

To quantitatively describe iron spectral features  we use a  model consisting of a power-law with exponential decay (\textit{cutoffpl} in {\sc xspec} package), a gaussian emission line  (\textit{gauss}) and  the absorption edge (\textit{edge}), {\sc xspec} formula \textit{edge*(cutoffpl+gauss)}. The line centroid and the energy of the edge were fixed at 6.4 keV and 7.112 keV, respectively. The broad band spectral shape of \obj\, is rather complex and to facilitate the description of the continuum with a simple model with a minimal number of free parameters, we decided to limit the energy range for spectral analysis to $3\textrm{--}12$ keV. 

Making the line centroid energy a free parameter of the fit (model \textit{cutoffpl+gauss}), its best-fitting values varied in the $\approx 6.35\textrm{--}6.45$ keV range. From our experience, this is within the accuracy of the PCA instrument in measuring the centroid energy for the iron line.  On the other hand, fixing the line energy at 6.4 keV and making the energy of the edge a free parameter of the model, we obtained the edge energy consistent with the value of 7.1 keV in all observations. This justifies our choice to freeze these parameters at values expected for the neutral iron.

In  HMXBs,  emission lines of Fe  XXV  and  Fe  XXVI are often observed \citep[][]{Aftab2019}, which can not be resolved at the PCA energy resolution.  Unfortunately, no useful CCD data is available for this outburst of \obj. An XMM-Newton observations during 2015 outburst is quite heavily piled up. However, the consistently good agreement of the line centroid and position of the K-edge  with the values expected for neutral iron in all analysed PCA data suggests that contribution of heavily ionized iron is not significant in this source.
 
 Equivalent hydrogen column density in this source was measured to be of the order of $N_H\approx\,10^{22}$ cm$^{-2}$ \citep{Tsygankov2016, Doroshenko2017}. Fixing $N_H$ at this value   in the spectral fitting  lead to poor fit quality with significant deviations  at lower energies. Making $N_H$ a free parameter of the fit results in the best-fitting values in the $\sim (0.2\textrm{--}0.5)\times 10^{22}$  cm$^{-2}$ range with some  variations throughout the outburst. Although some local absorption by circumbinary material is, in principle, possible in a binary system like \obj, insufficient low energy coverage of PCA instrument  and systematic uncertainties in its  energy response at the low energy end do not permit us to make any reliable statement about the value and time evolution of $N_H$. Therefore we  chose to exclude interstellar absorption  from the spectral model. This does not affect the best-fitting parameters of the continuum model in any significant way.   However, absorbing column with  $N_H\sim 10^{22}$ cm$^{-2}$ and cosmic abundance of iron should produce the iron K-edge with the optical depth of $\sim0.007\textrm{--}0.015$, with the exact value depending on the  cross-section models and assumed  abundance on iron along the line of sight in the direction toward \obj. This value is comparable, within a factor of $\sim 2\textrm{--}3$, with the measured value of the K-edge depth.  However, we clearly see variations of the iron K-edge depth, long-term, throughout the outburst, and short-term, on the time-scales corresponding to the rotation  period of the neutron star (Section \ref{par:phase_resolved}). This leaves no doubts that  most of iron K-edge absorption observed in the spectrum of the \obj\ originates in the source itself and not in the interstellar medium (ISM). 

In our fits, the spectral width of the line was poorly constrained but had a value typically $\sim0.3\pm0.1$ keV.  Therefore we fixed the width at 0.3 keV, a value similar to the results of \citet{Caballero2016,Baum2017}. With applied systematic errors (sect. \ref{par:data_analysis}) majority of  spectra have acceptable reduced chi-square value (the mean value  of $\approx 1.1$ for 16 dof) However, a few spectra have large values of the reduced chi-square, sometimes exceeding  $\sim 2$. Inspection of these spectra showed that large residuals are observed at the energies below $\la 5$ keV. The deviations of data from the model have  positive as well as negative sign and can not be universally  fixed varying the low energy absorption.  Based on our experience of working with PCA data we believe that, most likely, they are  related to some calibration uncertainties of the PCA instrument. 

Throughout the paper, all confidence intervals for spectral parameters are calculated for 68\% ($1\sigma$) confidence level (cl), and upper limits are 90\% cl. Table \ref{tab:rxte-spectra-table} presents the observation log, configuration and grouping used in phase-resolved spectroscopy, and results of the observation-averaged spectral analysis -- reduced chi-square value for the model described above, the $3\textrm{--}12$ keV model flux and the value of the equivalent width of the iron line, as well as its intensity.

\begin{figure*}
\centering
    \includegraphics[width=\textwidth]{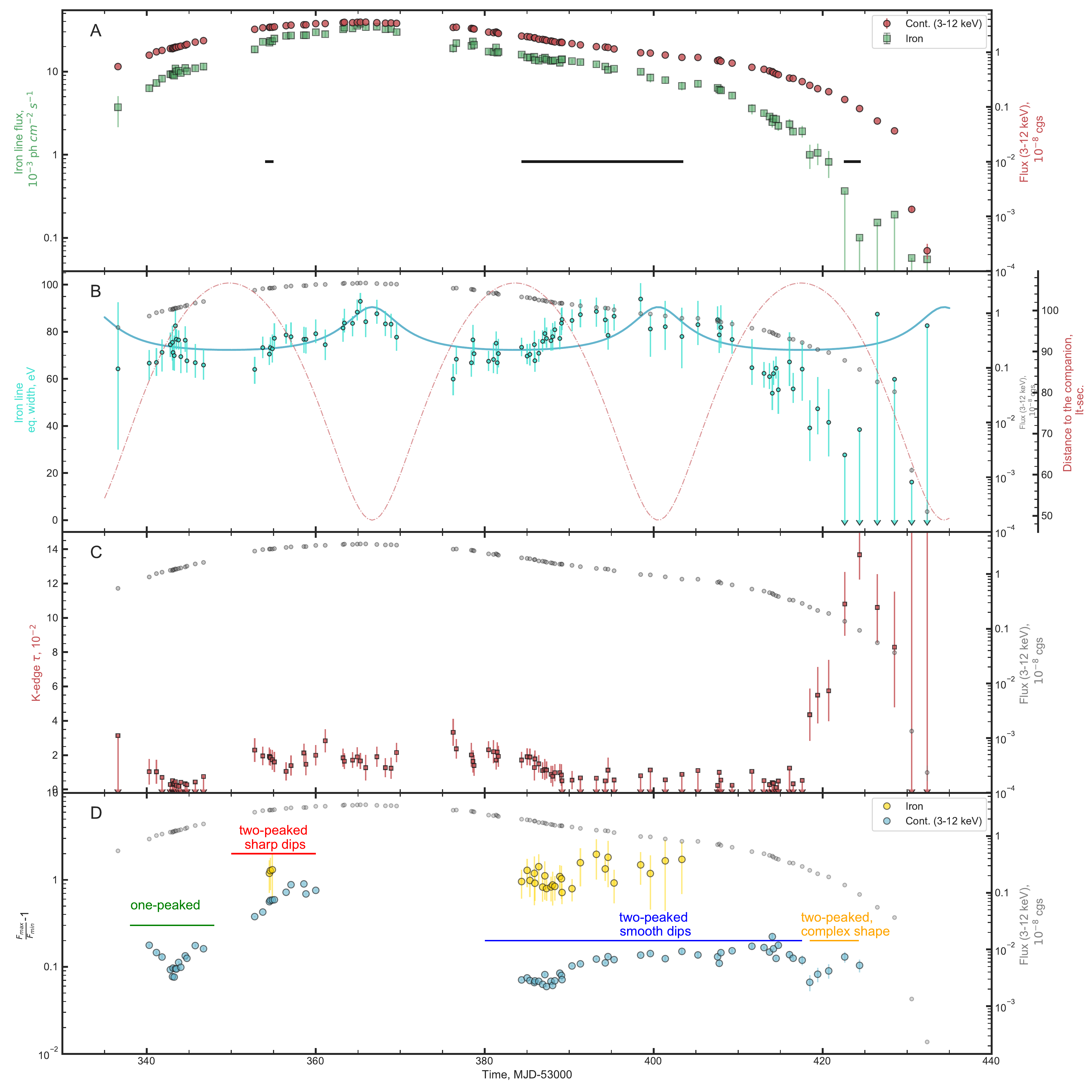}
    \caption{\textbf{Panel A}: The light curve of \obj\, in 2004-2005 outburst. The $3\textrm{--}12$ keV flux in units of $10^{-8}$ \ergpspcm\, is shown as red circles, while the flux of the \ironka\, line in units of $10^{-3}$  phot cm$^{-2}$ s$^{-1}$ is presented by green squares. Horizontal black solid lines indicate the observations during which phase-resolved spectroscopy was carried out. \textbf{Panel B}: The evolution with time of the equivalent width of the \ironka \, line (cyan circles). The line-of-sight projected distance between the pulsar and the optical companion is plotted with red dashed-dotted line. Solid cyan line shows the approximation of EW data with the model described by eq. \ref{eq:eqw_star}, see sect. \ref{par:discussion} for discussion. \textbf{Panel C}: Optical depth of the iron K-edge (red squares).  \textbf{Panel D}: Morphology of pulse profiles in the 3--12 keV  energy band. Cyan dots demonstrate the quantity $R=(F_{\rm max}-F_{\rm min})/F_{\rm min}$ for the 3--12 keV pulse profile, and yellow dots show the same but for iron line flux (see sect. \ref{par:phase_resolved}). Horizontal lines with text present approximate  classification of pulse profiles into morphological types (see the description in sect. \ref{par:pulse_prof}).
    For reference,  the $3\textrm{--}12$ keV flux in units $10^{-8}$ \ergpspcm\, is shown by  gray circles in panels B,C,D. }
    \label{fig:rxte-panno-ave_spe}
\end{figure*}

Fig. \ref{fig:rxte-panno-ave_spe}, panel A shows the evolution of the $3\textrm{--}12$ keV flux and  the iron line flux during the outburst. The outburst started approximately on MJD 53340 (2004 December 1), achieved its peak flux  of $3.5\times 10^{-8}$ \ergpspcm\, about 20 days later and then started its gradual decline followed by a fairly sharp drop in the end of the outburst around MJD 53430, after which observations of the source were finished. In the end of the outburst, the iron line flux declined faster than the continuum flux,  and the line was undetectable in the last six observations.

In Fig. \ref{fig:rxte-panno-ave_spe}, panel B the variation of the equivalent width of the iron line with time during the outburst is shown.  At the beginning of the outburst, the equivalent width had a value of $\sim65$ eV and started to grow until it reached $\sim90$ eV at the peak of the outburst. After that, the value of the equivalent width declined to $\sim70$ eV  until it grew again up to $\sim80$ eV in $\sim10$ days. 
 At the end of the outburst, the value returned to the initial value of $\sim$50\textrm{--}60 eV. There is a 'cut-off' in the end of the outburst, where the equivalent width drops (EW$<60$ eV) in a matter of $\sim5$ days. As the     equivalent width of the line depends  on its assumed spectral width  (in our case fixed at 0.3 keV), we also obtained spectral fits for smaller (0.2 keV)  and larger  (0.4 keV) line width. We found, as expected,  that although this changes slightly  the absolute value of the line EW,  the overall shape of the EW curve is not affected by the moderate changes of the line width. 

In Fig. \ref{fig:rxte-panno-ave_spe}, panel B one can see a hint of periodic variations  of the line EW with the period of $\approx 30$ days, consistent with the orbital period of the binary system. This behaviour was earlier reported by   \citet{Tsygankov2010b}. It could be related to the orbital motion of the system if, for instance, the reprocessing matter was located near the optical companion of the pulsar. In Fig. \ref{fig:rxte-panno-ave_spe}, panel B we plot by the dashed  line  the line-of-sight projected distance from the neutron star to the centre of the companion star. As one can see, the peaks of the line EW roughly coincide with the moments in time when the distance to the companion star is minimal. This suggests that some fraction of the iron line originates on the surface of the donor star or in the material located near it. 
This will be further discussed in Section  \ref{par:discussion}.

We report for the first time variability of the depth of the iron K-edge with time (Fig.~\ref{fig:rxte-panno-ave_spe}, panel C). At the beginning of the outburst, the iron K-edge was not detected in individual observations with the 90\% upper limits typically in the $\sim (0.5\textrm{--}1)\times 10^{-2}$  range. Around the peak of the outburst, the K-edge optical depth  was approximately constant at the level of $\approx 0.015\textrm{--}0.02$, and started to decline after MJD 53380 becoming undetectable again after MJD 53390 with the upper limit of $\sim (0.5-1)\times 10^{-2}$. However, after $\sim$  MJD 53420, approximately after apastron passage,  the optical depth of the K-edge increased significantly within about five days  to $\sim10\times 10^{-2}$ level  and then appeared to drop again, although the upper limits are not constraining to fully characterise its behaviour. Interestingly, the iron line was not detected when the K-edge was the deepest.  We note that the K-edge depth expected from the interstellar absorption appears to be somewhat higher than the upper limits shown in Fig.\ref{fig:rxte-panno-ave_spe}, panel C. This disagreement may be a result of  calibration uncertainties of \pca\ or related to deviations of the iron abundance from the adopted value, as discussed earlier in this section.

\begin{figure*}
\centering
    \includegraphics[width=\textwidth]{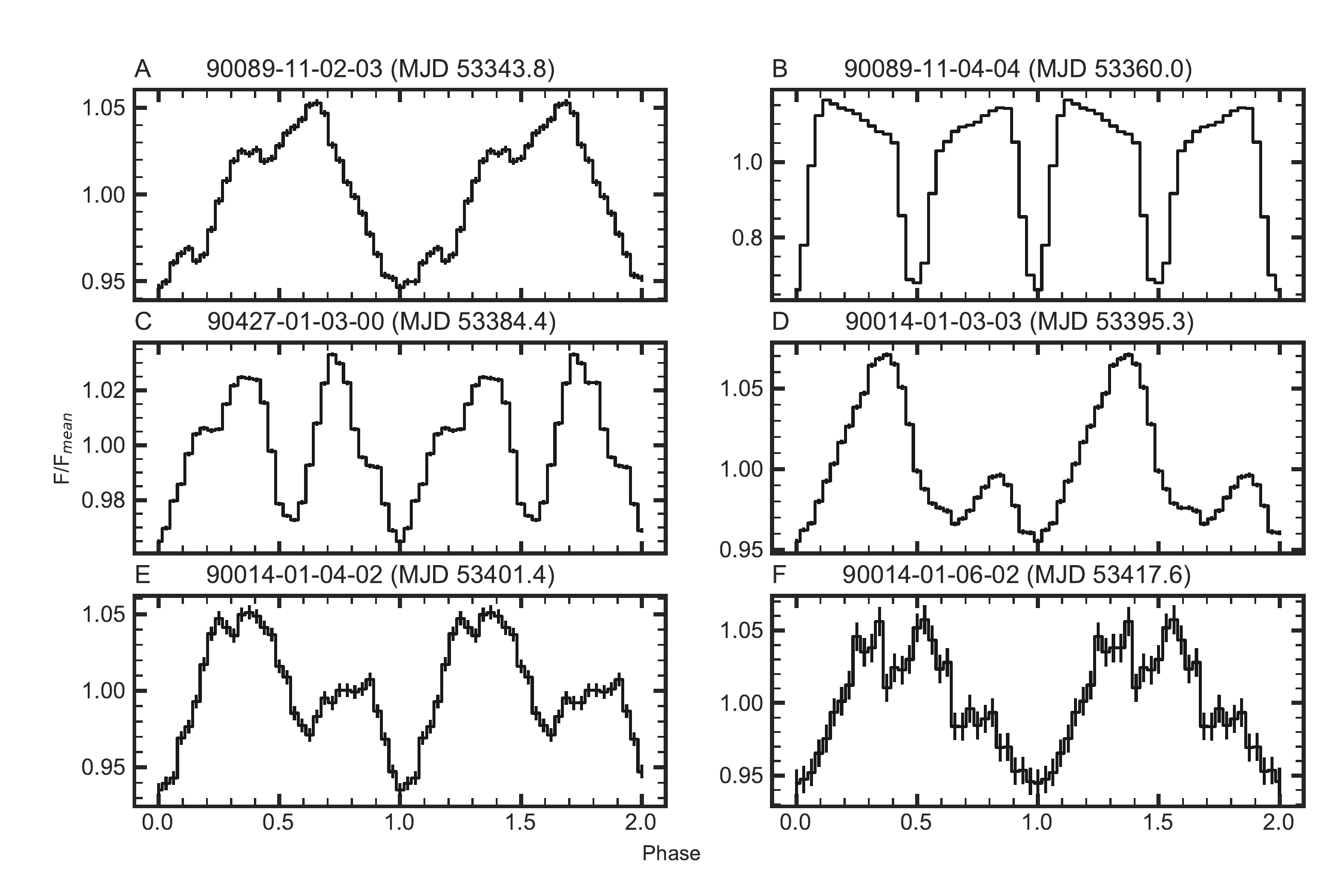}
    \caption{Pulse profiles in the 3--12 keV energy range at different stages of the outburst. Panel A shows a single-peaked pulse profile in the beginning of the outburst. The pulse profile shown in panel B was obtained near the peak  and displays two strong and narrow dips. Panels C, D and E show the  subsequent evolution of the pulse profile during  the declining phase of the outburst. Finally, in the panel F, near the end of the outburst, the pulse profile had a complex shape, seemingly double-peaked. The individual pulse profiles are not synchronised in phase, but shifted to have the main minumum at phase 0.}
    \label{fig:rxte-panno-atlas}
\end{figure*}

\subsection{Evolution of the pulse profile}\label{par:pulse_prof}

For each observation with sufficient time resolution (configurations B\_16ms\_64M\_0\_249. B\_16ms\_46M\_0\_49\_H, E\_125us\_64M\_0\_1s) we measured the pulsation period and obtained binary-motion corrected pulse profiles of counts rate in $3\textrm{--}12$ keV energy band \footnote{All pulse profiles, phase-resolved spectra  best-fitting parameters (including photon index, cutoff energy and reduced chi-square value) can be found at  \url{https://github.com/SergeiDBykov/v0332p53_materials} as well as in the online materials.}.

In the description below we use luminosities measured in the broad energy range 3-100 keV in \cite{Tsygankov2010, Lutovinov2015}, but assuming 5 kpc distance.

The evolution of the shape of the pulse profile followed the pattern typical for transient accreting X-ray pulsars, with complex evolution of pulse profiles in time as described below. Such behaviour has been reported in several pulsars \citep{Tsygankov2010,Koliopanos2016, Tsygankov2018,Wilson-Hodge2018,Lutovinov2021}  and predicted theoretically as a result of the switch of the emission diagram from a pencil beam to fan-shaped beam due to the changes in accretion regime \citep{Gnedin1973, Basko1976,  Mushtukov2015b}, although the detailed picture is still debated \citep[e.g.][]{Mushtukov2018}. 

At the beginning of the outburst (MJD 53340-50),  single-peaked $3\textrm{--}12$ keV pulse profiles were observed in virtually every observation. A representative pulse profile of this type is shown in Fig. \ref{fig:rxte-panno-atlas}, panel A. The pulse profile in this energy range had  one broad and somewhat asymmetric peak followed by a similarly broad smooth minimum,  the pulsed fraction\footnote{pulsed fraction, $PF$, defined as $PF=\frac{F_{\rm max}-F_{\rm min}}{F_{\rm  max}+F_{\rm min}}$ where $F_{\rm  max/min}$ is the maximum/minimum value of flux found in a pulse profile.} was $\sim5\%$. In \citet{Tsygankov2010} it was shown that the pulse profiles in \obj\, depend on the energy range, and in harder channels  may become double-peaked (e.g. $8\textrm{--}14$ keV profile at the rising phase of the outburst is double-peaked). Pulse profile also changes dramatically near the cyclotron line energy \citep[fig. 6]{Tsygankov2010}. These aspects of \obj\, have been a subject of detailed studies previously \citep{Tsygankov2006,Tsygankov2010} and not repeated here.

The transition to the double-peaked profile occurred during the data gap between MJD 53345-53350 at the luminosity level of $\sim 10^{38}$ \ergps.  Near the maximum of outburst, at MJD 53350 -- 53360, double-peaked profiles had two strong narrow dips and flat maxima, with the pulsed fraction of nearly $\sim 30\%$, separated in phase by $\Delta\phi\approx 0.5$ (Fig. \ref{fig:rxte-panno-atlas}, panel B). Interestingly, in two observations   90089-11-04-(03,04) (MJD 53358.8 and 53360.0) pulse profiles had the two dips with the same depth which may suggest some saturation or a full eclipse of the accretion column and the neutron star (see discussion in Section \ref{par:discussion}).  The peaks were also fairly equal in flux. Later in the outburst, after the timing data gap in MJD 53360--53385,  the breadth of the dips increased and the symmetry between the two peaks (and dips) disappeared (Fig. \ref{fig:rxte-panno-atlas}, panels C,D,E). The pulsed fraction returned to the initial $\sim5\%$. The shape of the pulse profile evolved in a complicated manner with its complexity  increasing (e.g. Fig. \ref{fig:rxte-panno-atlas}, panel F) towards  the end of the outburst, until luminosities as small as $\sim 1/30$ of the peak luminosity. As we did not detect the transition from double-peaked pulse profile to the singe-peaked one in the end of the outburst, we can conclude that the singe->double-peaked pulse profile transitions happen at very different luminosities in the beginning and the end of the outburst (hysteresis behaviour). 

We summarize the main features of the pulse profile evolution in Fig. \ref{fig:rxte-panno-ave_spe}, panel D where show the behaviour of the modulation amplitude $R=F_{ \rm max}/F_{ \rm min}-1$ and describe the morphology of the pulse profile at different stages of the outburst.

\begin{figure*}
\centering
    \includegraphics[width=\textwidth]{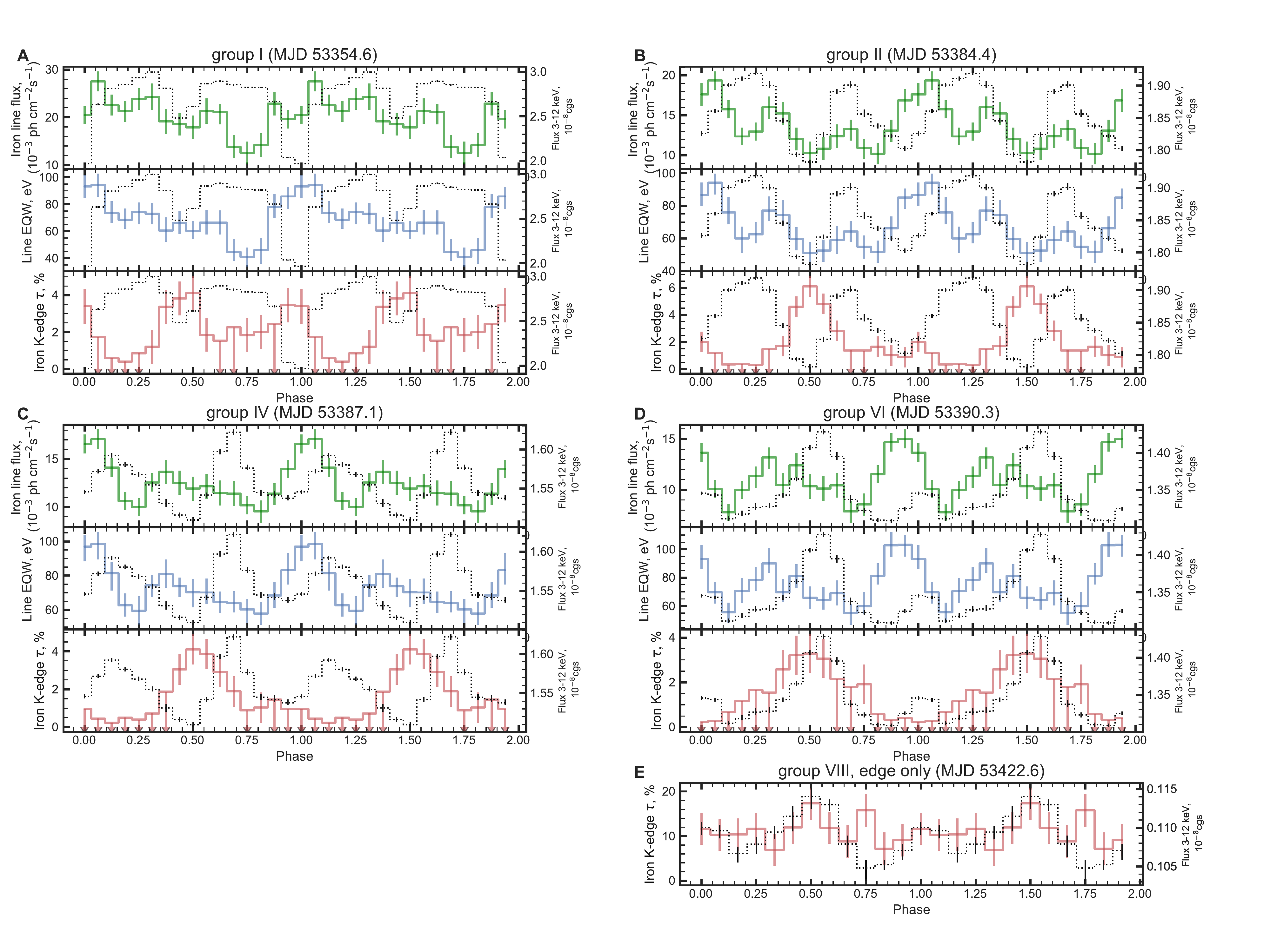}
    \caption{Pulse phase dependence of  parameters of the phase-resolved spectra for different groups of observations. See Table \ref{tab:rxte-spectra-table} for definition of data groups.  On each panel (A,B,C,D) we plot the pulse profile of the flux ($3\textrm{--}12$ keV) with a dotted black histogram on each sub-panel, the flux of the iron line (green histograms, top sub-panel) and its equivalent width (blue histograms, lower sub-panel), and the optical depth of the iron K-edge as a function of phase (red histograms, bottom sub-panel). \textbf{Panel A} shows the group I, the near-maximum part of the outburst (MJD 53350-60). \textbf{Panel B, C, D} show group II, IV and VI from the declining part of the outburst (MJD >53380). In \textbf{Panel E} we show the pulse profile of the total flux and K-edge depth  for the  group VIII, where no iron line was detected in the spectra.  All profiles were shifted along x-axis so that the maximum of K-edge optical depth is at phase 0.5.}
    \label{fig:rxte-panno-atlas-ph-res}
\end{figure*}

\subsection{Phase-resolved Spectroscopy}\label{par:phase_resolved}

Only 28  PCA observations were performed in data modes having sufficient time resolution and number of energy channels to permit pulse-phase-resolved spectroscopy of the iron line.  In Fig. \ref{fig:rxte-panno-ave_spe}, panel A the time  intervals covered by these observations are marked by two left-most black horizontal lines. As one can see, no phase-resolved spectroscopy is possible at the peak of the outburst.

 The data were divided  into seven groups of observations with  similar pulse profiles and  the same configuration/detectors. The groups cover different parts of the outburst and show different behaviour of the pulse profile and parameters of the phase-resolved spectra (see Table \ref{tab:rxte-spectra-table} and Fig. 
 \ref{fig:rxte-panno-atlas-ph-res}).  The spectrum in each phase bin was fitted in the 3--12 keV band with the model  {\it edge*(cutoffpl+gauss)} in {\sc xspec}. As suggested by the results of the  spectral analysis of  average spectra, we fixed the line energy and width at values of 6.4 keV and 0.3 keV respectively, and the edge energy at 7.1 keV (sect. \ref{par:phase_ave}). It should be noted that, unlike fluorescent line energy, the energy of the K-edge  depends notably on the ionization state even for weakly ionized iron (for example increasing from  7.1 keV for Fe I to 7.6 keV for Fe V), however, the higher edge energies seem to be excluded by the results of analysis of average spectra presented in Section \ref{par:phase_ave}.

Results of phase-resolved spectroscopy  are shown in Fig. \ref{fig:rxte-panno-atlas-ph-res}  where we present variations of  parameters of interest with the pulse phase.   In particular, we follow a total 3--12 keV flux, shown in each panel by the dashed histogram, flux and equivalent width of the 6.4 keV iron line and the optical depth of the iron K-edge (shown by solid histograms in the top, middle and bottom sub-panels respectively).

One of the most interesting findings of this work is the detection of the variable iron line K-edge whose optical depth varies with the pulse phase. It reveals itself most graphically in  data group I. This data group is the closest to the peak of the pulsar light curve in our sample, having the largest luminosity ($\sim 1.7 \times 10^{38}$ \ergps) among the data sets suitable  for phase-resolved spectroscopy.  The pulse profile on this date (around MJD 53354.6) shows two rather narrow dips of large (and unequal) amplitude separated by $\approx 0.5$ in pulse phase. In the deeper one of the two, observed flux from the pulsar drops by a factor of $\sim 1.5$.  Comparing with the collection of pulse profiles shown in Fig. \ref{fig:rxte-panno-atlas}  one may conclude that  with further increase of the luminosity,  about 5 days later, on MJD 53360,  the dips became more narrow and equalized in depth. The luminosity at that time was $\sim 15\% $ larger than in group I.  Notably, the dips in the pulse profile appear to be accompanied  by the peaks in the iron K-edge (Fig. \ref{fig:rxte-panno-atlas-ph-res}, panel A), suggesting that they may be caused by obscuration of the neutron star by, for example,  the accretion flow from the inner disc. This possibility is discussed in detail in the next section.  

Unfortunately, data around the maximum of the light curve were taken in  instrument configurations without 3--12 keV energy coverage. The next (in time) data set suitable for phase-resolved spectroscopy (group II, Fig. \ref{fig:rxte-panno-atlas-ph-res}, panel B) was taken a month later, on 
MJD 53384, when the luminosity was by a factor of $\sim 1.6$ lower than at the maximum of the light curve and by a factor of $\sim 1.5$ lower than in the previous data group  discussed above (group I). However, it still shows a clear peak in the K-edge depth  profile coinciding with one of the two dips in the total flux pulse profile. This trend continues in the following observations, data groups III, IV and V, albeit peaks and dips become broader and  progressively more smeared. However,  in the last data set (MJD 53390.3, group VI, Fig. \ref{fig:rxte-panno-atlas-ph-res}, panel D)  taken near the end of the outburst at the luminosity level of $\sim 0.09\times10^{38}$ \ergps, the pattern changes dramatically and the peaks of the K-edge absorption depth now coincide with the peaks of the total flux. Interestingly, at this point, the K-edge becomes undetectable in the average spectrum.

The iron line flux shows large variations with the pulsed fraction in the $\sim 30\%$ range without evident trends in time and luminosity, see Fig. \ref{fig:rxte-panno-atlas-ph-res}.  The pulse-profile shape of the line flux has a rather complex shape with several minima and maxima, and with less obvious relation to the modulation of the total flux. The equivalent width generally follows the iron line flux because the variability of the line is much larger than the variability of continuum emission, with a typical value of EW of $\sim 50\textrm{--}100$ eV. A remarkable property of the iron line spectral features is the lack of correlation between the iron line flux and the depth of its K-edge. This behaviour has important implications on the geometry of the neutral material producing these features which are  discussed in sect. \ref{par:discussion}.

After MJD 53420, pulsations were detected  only in two observations (90014-01-07-04 and 90014-01-07-00). For phase-resolved spectroscopy, we grouped these two observations into  data group VIII. As  the iron line was not detected in these data at all, we used the spectral model {\it cutoffpl*edge} which permitted us to  carry out phase-resolved spectroscopy  despite the missing channel at the energy 6.4 keV in the PCA instrument configuration E\_125us\_64M\_0\_1s used in these observations. 

We did not detect pulsations of the K-edge optical depth (Fig. \ref{fig:rxte-panno-atlas-ph-res}, panel E) in these data. One can see that the behaviour of the K-edge depth pulse profile differs drastically from that observed in previous data groups. In particular, the null hypothesis of the constant depth has  a  fully acceptable $\chi^2$ value of 7.8 for 11 dof. Assuming a sinusoidal pulse profile with the mean value equal to the K-edge optical depth $\tau_K$ measured in the average spectrum in these observations ($\approx 10\%$)  we obtained the $90\%$ confidence upper limit on the amplitude of $\tau_K$ variations of $\approx 60\%$ of the mean value. Due to the rather limited statistical quality of these data, the upper limit does not seem to be very constraining. However, $\tau_K$ pulsations with the pulse profile similar to those observed in data group II or IV can be excluded with high confidence. Indeed, assuming the pulse profile as in  panel B in Fig. \ref{fig:rxte-panno-atlas-ph-res} and re-scaling it to give the observed value of the mean K-edge optical depth in group VIII, we obtain  the $\chi^2=113$ (11 dof) for the null hypothesis of constant $\tau_K$.

\begin{figure}
    \centering
    \includegraphics[width=0.5\textwidth]{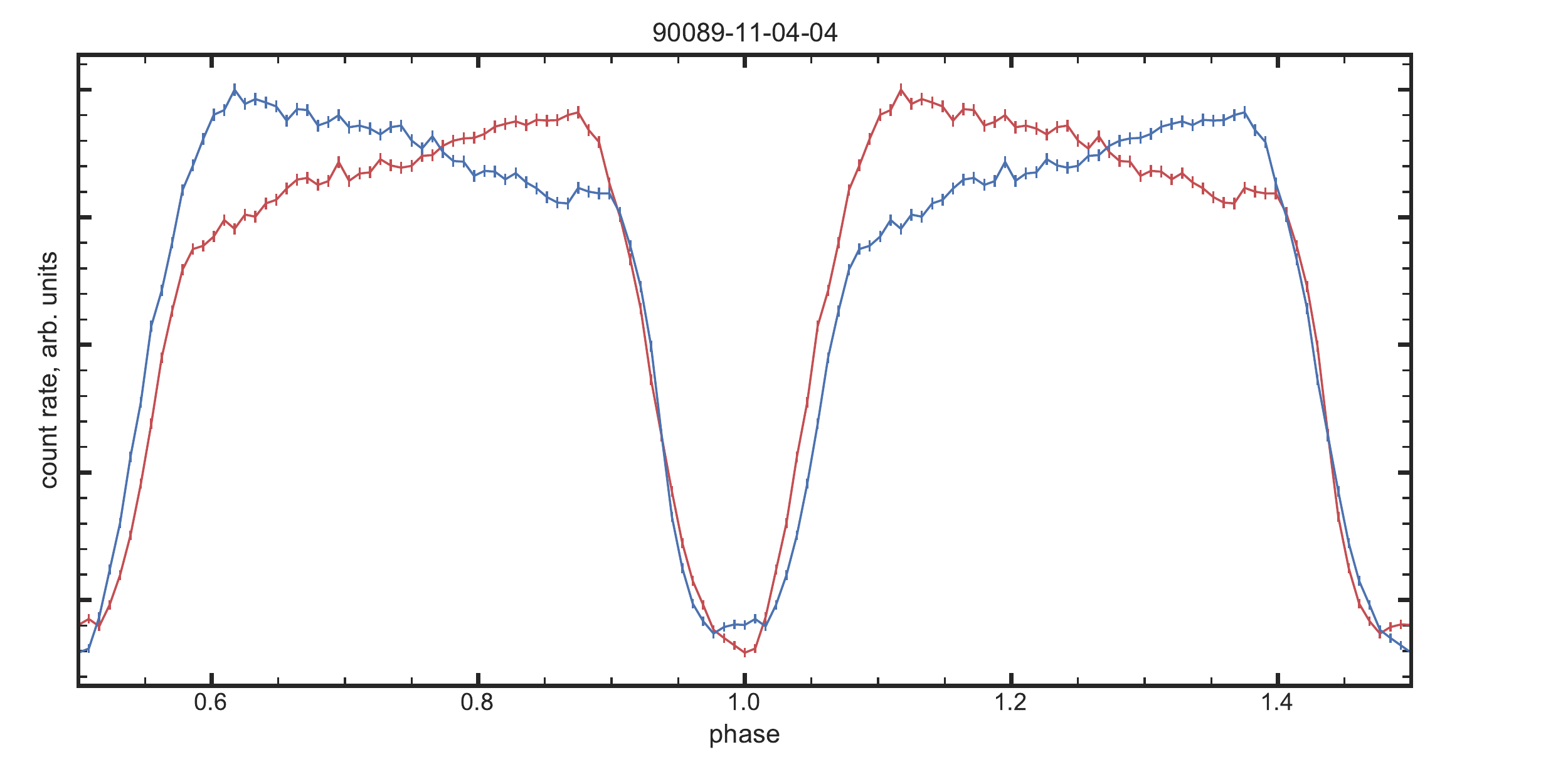}
    \caption{Count rate pulse profile in the 3--12 keV range for observation 90089-11-04-04 having maximum luminosity in our data (for which such a profile could be constructed) plotted in 128 phase bins (temporal resolution $\sim0.035$ sec). The units along y-axis are arbitrary. The blue line is the pulse profile, and red line is the same profile shifted by 0.5 in phase. This plot is intended to demonstrate the striking similarity between the shape and depth of the two dips and the fact that they are separated by nearly exactly $\Delta\phi=0.5$ in phase. }
    \label{fig:ph-res-pulse-profile}
\end{figure}

\begin{figure}
    \centering
    \includegraphics[width=0.5\textwidth]{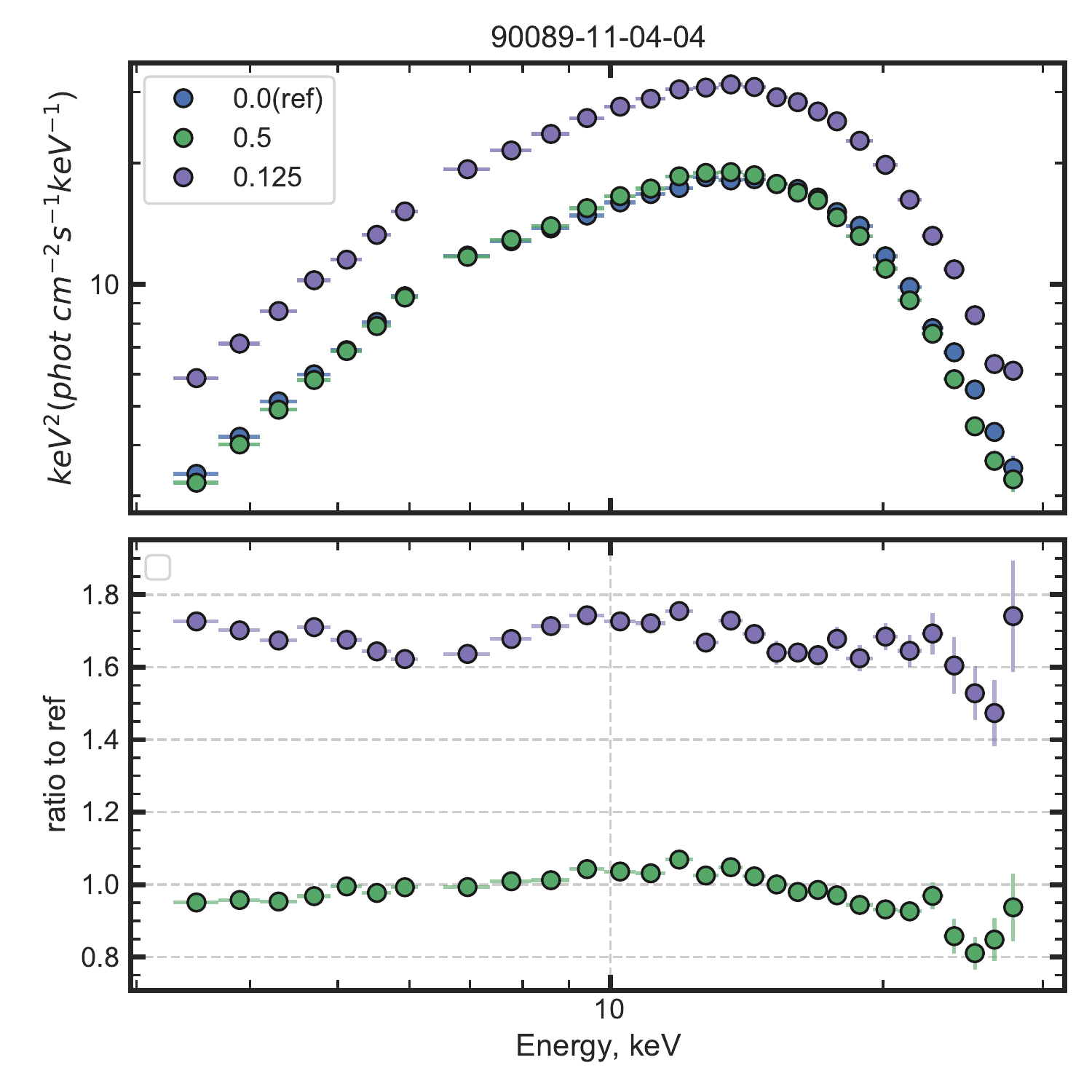}
    \caption{Top panel: Unfolded (with \textit{cutoffpl} model) spectra of three phase bins of the pulse profile shown in panel B in Fig. \ref{fig:rxte-panno-atlas} and in Fig. \ref{fig:ph-res-pulse-profile},  observation 90089-11-04-04. The chosen phase bins (specified in the legend) correspond to the bottom of the two dips (blue and green circles, phase bins 0/16 and 8/16) and the peak of the pulse profile (magenta circles, phase bin 2/16).  Bottom panel: the ratio to the phase 0 spectrum. The data gap at $\sim 6.5$ keV is due to configuration.}
    \label{fig:ph-res-ratio-dips}
\end{figure}

\section{Discussion}\label{par:discussion}

\subsection{Summary of observational picture}

The findings of this work can be summarized as follows:
\begin{enumerate}
    \item The 3--12 keV pulse profile evolves throughout the outburst from a single-peaked shape at low luminosity to a double-peaked one at high luminosity. The reverse transition was not observed at low luminosities in the end of the outburst. This behaviour is similar to other accreting transient X-ray pulsars in Be/X systems \citep{Tsygankov2007, Epili2017, Wilson-Hodge2018}, its features specific for \obj\ have been already extensively  discussed elsewhere, including its behaviour at higher energies \citep{Tsygankov2006}.
    \item Outside  the peak of the outburst, the pulse profiles in the 3--12 keV band show a moderate modulation at the level of $R=\left(F_{ \rm max}-F_{ \rm min}\right)/F_{ \rm min}\sim 0.1\textrm{--}0.2$. Near the peak of the outburst, the modulation reaches amplitude of $R\sim 1$ (Fig. \ref{fig:rxte-panno-ave_spe}, panel D). Qualitatively, this highly modulated pulse profile can be interpreted as the single peak profile similar to the one observed in the rising part of the light curve (panel A in Fig. \ref{fig:rxte-panno-atlas}), superimposed on which are two deep and narrow "absorption" features separated by 0.5 in pulse phase. In the declining part of the outburst, the pulse profiles have low modulation $\sim 0.1$ and double-peaked shape. Pulse profiles evolve in a complex manner after, possibly having a two-peaked shape until the end of the outburst. 
    \item The fluorescent iron line at the energy consistent with neutral or weakly ionized iron is detected throughout  most of the outburst. On the long time-scale, its equivalent width shows possible  modulation with the orbital period of the binary \citep[as was noted earlier in][]{Tsygankov2010b}, varying between $\approx 70\textrm{--}90$ eV. Notably, there is a clear cut-off in the end of the outburst, where the equivalent width drops down to $\la 60$ eV on the time-scale of about $\la 5\textrm{--}10$ days.
    \item The iron line flux pulsates with the rotational period of the pulsar, varying with the amplitude of $R\sim 1$, much larger than oscillations of the total flux. The pulsation amplitude of the iron line flux remains constant through the outburst, where phase-resolved spectroscopy was possible.
    \item We detected iron K-edge at the energy 7.1 keV with the optical depth significantly larger than expected from ISM absorption. Its optical depth varies with the pulse phase in a manner approximately anti-correlated with the  total flux and also evolves through the outburst.  Its long-term variation and presence of pulsations prove that it is mostly caused by the material near the neutron star itself.
    The presence of pulsating iron K-edge in \obj\ was not reported before.
\end{enumerate}

\subsection{Dips in the pulse profile}

The striking feature of the pulse profile at the peak of the outburst is the presence of two nearly identical dips separated in phase by exactly 0.5 (panel B in Fig. \ref{fig:rxte-panno-atlas} and Fig. \ref{fig:ph-res-pulse-profile} ). The source spectrum in the dips is (i) nearly identical to the spectrum outside the dips and (ii) the spectra in the two dips are nearly identical to each other in shape and  normalization (Fig. \ref{fig:ph-res-pulse-profile} and \ref{fig:ph-res-ratio-dips}). These properties and the narrowness of the dips suggest that they  could be caused by shadowing of the (part of) emission region by some scattering (i.e. ionized) material of moderate optical thickness. They can also arise due to shadowing of the part of the emission region by an opaque material. However, similarity of dip and off-dip spectra and, in particular, lack of any strong absorption features in the dip spectra, suggest that the dips can not be produced by neutral or weakly ionized absorbing material of moderate thickness. Similar pulse profiles with very sharp, luminosity-dependent dip-like features were observed earlier in several  accreting  X-ray pulsars such as GRO J1008-57 \citep{Naik2011}, EXO 2030+375 \citep{Epili2017},  A0535+262 \citep{Naik2008,Jaisawal2021},  GX 304-1 \citep{Jaisawal2016}, 2S 1417-624 \citep{Gupta2018}, 4U 1909+07  \citep{Jaisawal2020} and generally associated with the additional absorption of radiation by matter. 

The primary candidate for the obscuring material could be  the flow of matter from the inner part of the disc to the pulsar's surface (which is also called accretion curtain, or accretion channel). Indeed, from the continuity equation one can estimate the electron density $n_e$ in the accretion curtain  at radius $r$ as 
\begin{equation}
n_e(r)=\frac{\dot{M}/2}{v(r)S(r)\mu m_p}
\end{equation}

where $\dot{M}/2$ is the accretion rate through one accretion column assuming that there are two opposing streams as expected for the dipole configuration of the magnetic field, $v(r)$ is the velocity of the accreting matter (in the accretion channel  it should be close to the free-fall velocity, $v(r)\approx v_{\rm ff}(r) = \sqrt{2GM_{\rm NS}/r}$), $S(r)$ is the area of the flow, $\mu$ is the mean molecular weight, $n_e$ is the electron number density and  $m_p$ is the proton mass.
Expressing the distance from the neutron star to the part of the accretion flow causing obscuration as a fraction of the Alfven radius $r=ar_M$, one can estimate the area of the stream there as $S(r)=2\pi f d ar_M $, where $2\pi f$ is the azimuthal angular size  (in radians) of the accretion curtain, $d<<r$ is the thickness of the accretion curtain at this radius. With this one can express  the Thomson optical depth of the accretion curtain $\tau_T=n(r)\sigma_T d$ as follows:
\begin{equation}\label{eq:tau}
    \tau_T = \frac{1}{4\pi f a}\frac{\dot{M} \kappa_T}{r_M v_{\rm ff}(a r_M)} = 0.1 L_{38}^{8/7} R_6^{2/7} M_{NS}^{-10/7} B_{12}^{-2/7} f^{-1} a^{-1/2}  
\end{equation}
where $\kappa_T=\frac{\sigma_T}{\mu m_p}=0.34$ cm$^2$g$^{-1}$ is a electron scattering opacity for solar abundances, $\sigma_T$ is a electron scattering cross-section , and the scales are: $L_{38}$ for luminosity in units $10^{38}$ \ergps, $R_6$ for neutron star radius in units $10^6$ cm, $M_{NS}$ for neutron star mass in units of solar masses, $B_{12}$ for neutron star magnetic field in units $10^{12}$ G. The formula for Alfven radius was taken from \cite{Frank2002}.

Near the peak of the outburst,  the luminosity of \obj\ was $\sim1.9\times 10^{38}$ erg/s \citep[but assuming 5 kpc distance]{Lutovinov2015}; the azimuthal size of the accretion curtain can be estimated from the width of the dips in the pulse profile (panel B in Fig. \ref{fig:rxte-panno-atlas}) $f\sim 0.2$,  the strength of the magnetic filed on the surface was measured at  $B=2.5\times10^{12}$ G (see sect. \ref{par:intro})  and, assuming $a=0.5$ we  obtain 
\begin{equation}
    \tau_T\simeq 0.9
\end{equation}
 We note that this result is close to the one in  \citet[equation (3), $\lambda=0$ and $\tau_e = L_{39}^{8/7}B_{12}^{-2/7}$]{Mushtukov2017} if we take into account that the azimuthal angle of an accretion channel is not equal to $2\pi$.

Thus, the accretion curtain is sufficiently optically thick to explain the dips in the pulse profile with the amplitude $R\sim 1$, assuming that its material is (nearly) fully ionised.  However, the biggest difficulty of this scenario is the presence of two dips separated by almost exactly $\Delta \phi=0.5$, see Fig. \ref{fig:rxte-panno-atlas}, panel B. Assuming that the magnetic filed has the dipole geometry, we can not identify a configuration in which  two  accretion flows could produce two nearly  identical dips separated by $\Delta \phi=0.5$ in phase (Fig. \ref{fig:ph-res-pulse-profile}).

\citet{Mushtukov2018} proposed that the dips in the pulse profile at the highest luminosity are caused by the eclipse of the accretion column  by the neutron star itself.  They noticed the fact that radiation from the accretion column is strongly beamed towards the surface of the neutron star \citep{Poutanen2013, Mushtukov2018}. Therefore, during the dip observer receives mainly emission from the eclipsed accretion column on the opposite side, focused by the neutron star serving as a gravitational lens   \citep[fig. 6 in ][]{Lutovinov2015}. Such a model naturally explains  two dips separated in phase by $\Delta \phi=0.5$. However, in the simplified models considered in \citet{Mushtukov2018}, the dips in the pulse profile are not symmetric, having different depth. It remains to be seen whether a more accurate account for the emission diagram of the accretion column can improve this aspect of the model \citep{Mushtukov2018}. We also note that according to the estimations above, the accretion flow from the inner disc presents a substantial obstacle for radiation escaping from the immediate vicinity of a pulsar, even if its azimuthal expansion $f$ is larger than $0.2$.

\subsection{Pulsating   K-edge of iron}

The instrument configuration did not permit us to perform phase-resolved spectral analysis at the highest luminosities, where the dips were most pronounced. The nearest observation for which this is possible was carried out on around MJD 53354.6 about $\approx 6$ days earlier. The pulse profile and results of phase-resolved spectral analysis are shown in panel A of Fig. \ref{fig:rxte-panno-atlas-ph-res}. The two dips are clearly seen in the pulse profile, albeit of unequal depth and separated in phase by $\Delta \phi \ne 0.5$. Notably,  two clear K-edge absorption peaks coinciding with the dips in the total flux. The optical depth of the edge is  fairly small,  $\tau\approx 0.04$, therefore  they  can not be associated with the material producing the dips. Indeed, considering the accretion curtain as an example, we note that  for  the solar abundance of cosmic elements, the optical depth of the iron K-edge is $\tau_{\rm Fe-K}\sim2\tau_T$. Given our estimate of the Thomson optical depth of the accretion curtain above (eq. \ref{eq:tau}), $\tau_T\sim1$, this is grossly inconsistent with the observed depth of the K-edge in the phase-resolved spectrum. The K-edge of such small depth could result from an addition of a small fraction of emission reprocessed by neutral material of solar abundance, for example in the accretion disc or on the surface of the neutron star near the accretion column (in the latter case the gravitational redshift should be taken into account).  Any such model should also explain the anti-correlation between the total flux and the depth of the K-edge.  To add to the complexity of the observational picture, we mention that pulsating K-edge was detected in all analysed data groups after the peak of the light curve, with a rather complicated pulse profile. 
 The lack of correlation between the depth of the edge and flux or equivalent width of the  fluorescent iron line suggests that the line and the edge originate in different locations. Indeed, if, for example,  the reprocessing neutral material was distributed  symmetrically around the primary source, the optical depth of K-edge should have been correlated with the equivalent width of the fluorescent line. Obviously, our results exclude such simple geometries.

The surge in the  K-edge depth near the end  of the outburst (Fig. \ref{fig:rxte-panno-ave_spe}, panel C) is quite interesting. The lack of pulsations of the K-edge optical depth during this period suggests that the absorbing material is located sufficiently far from pulsar and, likely, has a different origin, than pulsating K-edge absorption observed in the earlier stages of the outburst. The duration of the K-edge surge, $\sim 10-15$ days, is about $\sim 1/3\textrm{--}1/2$ of the orbital period of the binary, suggesting that absorbing material may be associated with some circumbinary material, for example with wind from the donor star. The lack of the fluorescent line of iron in the spectrum during this period further supports the scenario that we are dealing with an absorbing screen. In principle, the spectra in this period may be fitted with a simple absorbed power-law model, with a rather large hydrogen column density, $N_H\sim\,(2.3 \textrm{--} 3) \times 10^{22}$ cm$^{-2}$. With this fit, some residual K-edge absorption was still detected in some of the observations. However, as discussed above, the limited low energy coverage  of \pca\, did not permit us to make a definitive conclusion. Such absorption enhancements are observed in different X-ray pulsars \citep[e.g.][]{Hemphill2014, Jaisawal2014,Sanjurjo-Ferin2021,Liu2021, Ji2021} and usually related to the structures like wind or accretion stream through the inner Lagrange point. To our knowledge, such a clear and isolated episode of increased absorption is detected in Be/X-ray binary system for the first time.

\subsection{6.4 keV line}

The modulation of the iron line equivalent width with the orbital period of the binary, first noted in \cite[][their fig. 2 ]{Tsygankov2010b}, and its tentative anti-correlation with the distance to the companion star (Fig. \ref{fig:rxte-panno-ave_spe}) suggest that  some fraction of the iron line flux originates on the surface of the donor star or in the neutral material in its vicinity. In this context we note that optical studies of \obj\ suggested the presence of a circumstellar (decretion) disc \citep{Negueruela1999, Caballero2016}. Although its extent in \obj\  is not known, based on typical parameters of such disks in other systems \citep[][]{Coe2015} we conclude that it can contribute to the fluorescent iron line emission  observed from this source.

To estimate the fraction of emission from donor star and its circumstellar disk we fit the equivalent width curve with the model 
\begin{eqnarray}\label{eq:eqw_star}
EW(t)=EW_0 + EW_{\rm star}(t)\\
EW_{\rm star}(t)=\frac{N_{star}}{D_{\rm star}^2(t)}
\end{eqnarray}
where $D_{\rm star}(t)$ is the distance from the neutron star to the optical companion, $EW_{\rm star}(t)$ is the equivalent width of the line produced at or near the optical companion and modulated with the orbital period of the binary and $EW_0$ represent the rest of the iron line flux.  $N_{\rm star}$ is a normalization factor. This procedure is similar to the one used in \cite{Tsygankov2010b}. During the fitting procedure we ignored the observations without iron line in spectra (i.e. after approximately MJD 53420). The model employed here ignores possible variations of the emission diagram of the pulsar throughout the outburst, assuming that $EW_{\rm star}(t)$ is determined only by the solid angle of the companion star as seen from the pulsar. This approximation is sufficient for the purpose of this estimate. Fitting to the data we found $EW_0=67\pm 2$ eV. The best-fitting model is shown by the solid line in Fig. \ref{fig:rxte-panno-ave_spe}, panel B. 

Thus, we find that   the equivalent width of the iron line from the  donor star varies between $EW_{\rm min}\approx 5$ and $EW_{\rm max}\approx 25$ eV. This is  consistent with the expectation \citep{Basko1974}. Indeed, the radius of the donor star in \obj\ is probably close to $\approx 21$ light sec (nine solar radii) \citep[][according to spectral type]{Negueruela1999}, the solid angle subtended by the donor star varies in the range $\Delta\Omega/4\pi\approx 0.01 \textrm{--} 0.044$. For an isotropic emission diagram of the primary  emission, the equivalent width of the fluorescent line is $EW_{\rm max} \times \Delta\Omega/4\pi$ with $EW_{\rm max} \sim 1$ keV \footnote{for  moderate optical depth  
$$EW_{\rm max}= 0.74 \frac{\int_{7.1}^{\infty} I(E)(7.1/E)^{2.8}dE}{I(E=6.4 keV)}\tau\ {\rm keV}$$
assuming roughly solar abundance \citep[][formula 2]{Churazov1998}, where $I(E) [phot\,s^{-1} cm^{-2} keV^{-1}]$ is the spectrum of a source.
For a power-law spectrum with photon index $\Gamma\sim -0.5$ and cutoff energy $E_{cut}=7$ keV (parameters typical for \obj) the above formula gives   $EW_{\rm max}\approx 1.8 \tau$ keV. We neglected the effect of CRSF on the flux above 7.1 keV.}  and will vary between $\sim 10$ and $\approx 44$ eV. The observed values are within a factor of $\sim 2$ lower which may be understood as the result of the the anisotropy of the emission diagram of the pulsar. he presence of the  circumstellar disk around the donor star will further complicate the picture. Furthermore, the fluorescent line flux depends on the inclination of the system which was not taken into account in our simple estimate.

Results of the phase-resolved spectroscopy show that the iron line flux pulsates with the rotation period of the neutron star (Fig. \ref{fig:rxte-panno-atlas-ph-res}). Due to the large size of the binary system, $\sim 10^2$ light sec, much larger than the rotation period of the neutron star, $\sim 4.4$ sec, the pulsating component of the iron line can not originate on (or near) the surface of the donor star. The complex pulse profiles of the line flux and equivalent width suggest that the pulsating part of the iron line  must be associated with  neutral material located within $\ll 1$ light sec from the neutron star. As estimated above, the total equivalent width of the fluorescent line originating near the neutron star can not exceed $\la 67$ eV. On the other hand, the amplitude of pulsations of the iron line equivalent width is $\sim 50\textrm{--}60$ eV. Thus, the pulsating component of the fluorescent iron line is very strongly modulated, varying by a factor of $\sim 5-10$ with the rotational phase of the pulsar. 

Such a strong modulation of the iron line flux seems to be difficult to explain, if the line was originating due to reflection off the accretion disc, because of its large solid angle as seen from the pulsar. 

A plausible fluorescence site in the vicinity of the neutron star is the accretion curtain. The solid angle subtended by the two  accretion flows as seen from the emission region equals $\Delta\Omega\sim 2\times \pi/2\times 2\pi f$, where, as before, $2\pi f$ the azimuthal angular size (in radians) of the accretion curtain. Thus, $\Delta\Omega/4\pi\sim \pi/2\times f\sim 0.3$, assuming $f\sim 0.2$. Taking into account that the accretion flow is moderately thick (see above), the maximum value of the equivalent width of the line it can produce is a few hundred eV, i.e. it can explain the observed equivalent width of the iron line. As the accretion curtain can have a rather large ratio of its sizes in the azimuthal and radial directions, its rotation can easily explain the large modulation of the iron line, although particular details on the light curve need a much more detailed consideration with the account for the geometry of the accretion flows and the emission diagram of the pulsar.

\section{Conclusions}\label{par:conclusion}

We analysed the data of \pca\ observations of \obj\ during its type II outburst in 2004. We paid particular attention to the  variability of the iron spectral features -- the fluorescent  line at 6.4 keV and K-edge at 7.1 keV and investigated their evolution  on the time-scale of the outburst and their pulsations with the pulsar period. 
Detection of the pulsating iron K-edge in \obj\ is reported for the first time in this paper.

 Both iron line and edge  show a complex dependence on the pulse phase which can not be self-consistently accommodated in any of the existing models of emission of accreting pulsars.
The most striking of these features are:
\begin{enumerate}
    \item at high luminosity the pulse profiles in the $3\textrm{--}12$ keV band have two deep and narrow dips of $F_{\rm max}/F_{\rm min} \sim 2$ and of nearly identical shape, separated in phase by  exactly 0.5. The source spectrum is identical during the two dips and very similar in shape to the spectrum outside the dips.
    \item the K-edge peaks at the phase intervals corresponding to the dips, however, the optical depth of the edge $\tau_K\sim 0.05$ is by far insufficient to explain the dips.
    \item after accounting for the contribution of reflection from the donor star and any circumstellar material which may be present near it,  the iron line flux shows pulsations of very large amplitude with the modulation of $F_{\rm max}/F_{\rm min} \sim 10$. The pulse profile of the line flux does not show any easily identifiable correlations with the pulse profiles of the total flux or K-edge depth.
\end{enumerate}

\obj\ presents rich and unique opportunities and motivation for further development of theoretical models of emission of accreting neutron stars.

\section*{Acknowledgements}
 We thank the anonymous referee for useful and constructive comments and suggestions which helped to improve the presentation of our results.
This work was supported by the grant of the Ministry of Science and Higher Education of the Russian Federation 14.W03.31.0021. SB acknowledges support from and participation in the International Max-Planck Research School (IMPRS) on Astrophysics at the Ludwig-Maximilians University of Munich (LMU). 

\section*{Data Availability Statement}
We are grateful for the \pca\ data to the High Energy Astrophysics Science Archive Research Center (HEASARC) Online Service \footnote{\url{https://heasarc.gsfc.nasa.gov/cgi-bin/W3Browse/w3browse.pl}} provided by the NASA/Goddard Space Flight Center. To download data of \rxte\, used in this work, one should use observations ID from Table \ref{tab:rxte-spectra-table}.



\bibliographystyle{mnras}
\bibliography{literature.bib} 

\newpage

\appendix

Table \ref{tab:rxte-spectra-table} shows an observation log of \pca\, data used in this work (Group number, Observation ID, exposure, time and configuration used in phase-resolved spectroscopy), and the main parameters of spectral fits: reduced chi-squared value (11 dof),  flux and equivalent width of the iron line, and flux in $3\textrm{--}12$ keV band, and the intensity of the iron line.

\section{\pca\, observation log}
\centering
\input{figures/rxte/allpars_edge_cutoffpl}

\onecolumn

\bsp	
\label{lastpage}
\end{document}

%% file: figures/rxte/allpars_edge_cutoffpl.tex
\onecolumn
\begin{longtable}{ccccccccc}
\caption{The log of  \pca\, observations of the 2004-2005 outburst of \obj.}\\
\label{tab:rxte-spectra-table}\\
\hline
 Gr. &            ObsID & \begin{tabular}[c]{@{}l@{}}Time\\ MJD\end{tabular} & \begin{tabular}[c]{@{}l@{}}Exposure\\ s\end{tabular} &            Configuration* & $\chi^2_{red}$ &     \begin{tabular}[c]{@{}l@{}}Eq. width\\ eV\end{tabular}    &   \begin{tabular}[c]{@{}l@{}}Flux (3-12 keV)\\$10^{-9}\times$ erg  cm$^{-2}$ s$^{-1}$\end{tabular}        & \begin{tabular}[c]{@{}l@{}}Iron line int\\ $10^{-3}$ phot cm$^{-2}$ s$^{-1}$ \end{tabular}  \\ \hline
    - &   90089-11-01-00 &   53336.6 &         160 &                       - &           1.06 &  $64^{+28}_{-34}$ &   $5.48^{+0.05}_{-0.03}$ &   $3.7^{+1.3}_{-1.6}$ \\
    - &   90089-11-01-02 &   53340.3 &        2688 &                       - &           0.93 &    $67^{+6}_{-7}$ &            $8.85\pm0.02$ &           $6.3\pm0.6$ \\
    - &   90089-11-01-03 &   53341.1 &        2912 &                       - &           0.78 &          $67\pm6$ &           $10.20\pm0.02$ &           $7.3\pm0.7$ \\
    - &   90089-11-01-04 &   53341.8 &        1760 &                       - &           0.54 &    $71^{+6}_{-8}$ &  $10.82^{+0.02}_{-0.01}$ &   $8.2^{+0.6}_{-0.8}$ \\
    - &   90089-11-02-00 &   53342.8 &       11584 &                       - &           1.49 &          $75\pm5$ &           $11.70\pm0.01$ &           $9.3\pm0.4$ \\
    - &   90089-11-02-05 &   53343.0 &        1712 &                       - &           1.52 &    $76^{+7}_{-8}$ &           $11.62\pm0.02$ &           $9.3\pm0.7$ \\
    - &   90089-11-02-06 &   53343.1 &        2288 &                       - &           1.51 &    $71^{+6}_{-7}$ &           $12.06\pm0.02$ &           $9.1\pm0.6$ \\
    - &   90089-11-02-01 &   53343.2 &        2000 &                       - &           1.43 &    $70^{+8}_{-6}$ &           $11.96\pm0.02$ &           $8.9\pm0.6$ \\
    - &   90089-11-02-02 &   53343.4 &        1424 &                       - &           1.88 &    $83^{+7}_{-6}$ &           $12.44\pm0.02$ &          $10.9\pm0.7$ \\
    - &  90089-11-02-03G &   53343.5 &       10400 &                       - &           1.86 &          $77\pm4$ &           $12.28\pm0.01$ &          $10.0\pm0.5$ \\
    - &   90089-11-02-03 &   53343.8 &       11680 &                       - &           2.07 &          $77\pm4$ &           $12.63\pm0.01$ &          $10.3\pm0.5$ \\
    - &   90089-11-02-07 &   53344.0 &        1824 &                       - &           1.51 &    $69^{+6}_{-7}$ &           $12.92\pm0.02$ &           $9.6\pm0.7$ \\
    - &   90089-11-02-04 &   53344.5 &        2816 &                       - &           1.48 &    $76^{+6}_{-5}$ &           $13.50\pm0.02$ &          $11.0\pm0.6$ \\
    - &   90089-11-02-10 &   53344.7 &        1936 &                       - &           2.19 &    $68^{+9}_{-5}$ &           $13.96\pm0.02$ &          $10.1\pm0.7$ \\
    - &   90089-11-02-09 &   53345.7 &        2000 &                       - &           2.59 &    $67^{+8}_{-6}$ &  $15.31^{+0.03}_{-0.02}$ &          $10.9\pm0.8$ \\
    - &   90089-11-02-08 &   53346.7 &        1408 &                       - &           0.66 &    $66^{+7}_{-6}$ &           $16.32\pm0.03$ &          $11.5\pm1.2$ \\
    - &   90089-11-03-03 &   53352.8 &        1152 &                       - &           1.15 &    $64^{+5}_{-6}$ &           $26.30\pm0.05$ &          $18.5\pm1.8$ \\
    - &   90089-11-03-04 &   53353.7 &        3168 &                       - &           0.87 &    $73^{+5}_{-4}$ &           $28.04\pm0.04$ &          $22.7\pm1.5$ \\
    I &  90089-11-03-00G &   53354.5 &        4416 &  X &           0.67 &    $71^{+4}_{-5}$ &           $28.60\pm0.04$ &          $22.3\pm1.5$ \\
    I &  90089-11-03-01G &   53354.6 &       10768 &  X &           0.48 &          $73\pm4$ &           $28.76\pm0.04$ &          $23.3\pm1.4$ \\
    I &   90089-11-03-02 &   53354.9 &        2128 &  X &           0.78 &          $73\pm6$ &           $28.68\pm0.04$ &          $23.1\pm1.6$ \\
    - &   90089-11-03-05 &   53355.1 &        1904 &                       - &           0.63 &    $77^{+6}_{-5}$ &           $29.21\pm0.04$ &          $25.0\pm1.7$ \\
    - &  90089-11-04-00G &   53356.5 &        3008 &                       - &           0.63 &          $79\pm4$ &           $30.68\pm0.04$ &          $26.8\pm1.6$ \\
    - &   90089-11-04-01 &   53357.1 &        1808 &                       - &           0.90 &    $78^{+6}_{-4}$ &           $31.40\pm0.05$ &          $27.2\pm1.8$ \\
    - &  90089-11-04-02G &   53358.6 &        2672 &                       - &           0.54 &    $77^{+5}_{-6}$ &           $31.83\pm0.04$ &          $27.3\pm1.7$ \\
    - &   90089-11-04-03 &   53358.8 &        1216 &                       - &           0.66 &    $77^{+5}_{-6}$ &           $31.82\pm0.05$ &          $27.3\pm2.0$ \\
    - &   90089-11-04-04 &   53360.0 &        1600 &                       - &           0.98 &    $79^{+6}_{-4}$ &           $33.42\pm0.05$ &          $29.6\pm2.0$ \\
    - &   90089-11-04-05 &   53361.1 &         944 &                       - &           0.76 &          $75\pm7$ &           $33.52\pm0.06$ &          $28.2\pm2.3$ \\
    - &  90089-11-05-00G &   53363.2 &        2768 &                       - &           0.89 &    $82^{+4}_{-6}$ &           $34.77\pm0.05$ &          $32.2\pm1.9$ \\
    - &  90089-22-01-00G &   53363.4 &       20544 &                       - &           0.39 &    $84^{+6}_{-4}$ &           $35.32\pm0.04$ &          $33.6\pm1.7$ \\
    - &  90089-22-01-01G &   53364.4 &       21792 &                       - &           0.58 &    $84^{+3}_{-5}$ &           $34.78\pm0.04$ &          $33.0\pm1.6$ \\
    - &   90089-11-05-01 &   53364.9 &        1968 &                       - &           0.65 &    $88^{+5}_{-4}$ &           $35.36\pm0.05$ &          $35.5\pm2.0$ \\
    - &  90089-11-05-08G &   53365.3 &       15392 &                       - &           0.59 &          $93\pm4$ &           $35.55\pm0.04$ &          $37.6\pm1.7$ \\
    - &   90089-11-05-02 &   53365.9 &         624 &                       - &           0.63 &    $84^{+6}_{-7}$ &           $35.59\pm0.07$ &          $34.2\pm2.7$ \\
    - &  90427-01-01-00G &   53367.2 &        1584 &                       - &           1.03 &    $88^{+6}_{-4}$ &           $34.78\pm0.05$ &          $34.6\pm2.1$ \\
    - &   90427-01-01-01 &   53368.2 &        2240 &                       - &           0.82 &          $83\pm5$ &           $34.10\pm0.05$ &          $32.3\pm1.9$ \\
    - &   90427-01-01-02 &   53368.9 &        1264 &                       - &           0.37 &    $83^{+4}_{-6}$ &           $34.38\pm0.05$ &          $32.6\pm2.2$ \\
    - &   90427-01-01-03 &   53369.6 &        1872 &                       - &           0.67 &          $78\pm6$ &           $33.84\pm0.05$ &          $29.8\pm2.0$ \\
    - &   90427-01-02-02 &   53376.3 &         688 &                       - &           1.39 &    $60^{+6}_{-7}$ &           $28.48\pm0.06$ &          $19.1\pm2.2$ \\
    - &   90427-01-02-03 &   53376.6 &        2240 &                       - &           0.94 &          $68\pm6$ &           $28.76\pm0.04$ &          $22.0\pm1.6$ \\
    - &   90014-01-01-00 &   53378.4 &        1040 &                       - &           0.98 &    $67^{+5}_{-6}$ &           $27.33\pm0.05$ &          $20.3\pm1.9$ \\
    - &   90014-01-01-06 &   53378.6 &        1344 &                       - &           0.54 &          $77\pm6$ &           $27.09\pm0.05$ &          $23.1\pm1.7$ \\
    - &   90014-01-01-07 &   53378.7 &        1744 &                       - &           0.76 &    $71^{+6}_{-5}$ &           $26.62\pm0.04$ &          $20.9\pm1.6$ \\
    - &   90014-01-01-03 &   53380.5 &        8144 &                       - &           0.76 &          $67\pm4$ &           $23.38\pm0.03$ &          $17.4\pm1.1$ \\
    - &   90014-01-01-02 &   53381.0 &        1376 &                       - &           1.23 &    $68^{+7}_{-6}$ &           $22.87\pm0.04$ &          $17.2\pm1.5$ \\
    - &   90014-01-01-01 &   53381.3 &        3280 &                       - &           1.12 &          $75\pm5$ &           $23.48\pm0.03$ &          $19.5\pm1.3$ \\
    - &   90014-01-01-04 &   53381.5 &        2784 &                       - &           0.52 &          $67\pm5$ &           $22.81\pm0.03$ &          $16.8\pm1.3$ \\
    - &   90014-01-01-05 &   53381.6 &        2176 &                       - &           0.54 &          $71\pm6$ &           $22.12\pm0.04$ &          $17.2\pm1.3$ \\
   II &   90427-01-03-00 &   53384.4 &       13152 &  Y &           0.66 &          $73\pm4$ &           $19.79\pm0.02$ &          $16.0\pm0.9$ \\
   II &   90427-01-03-01 &   53385.0 &        9776 &  Y &           0.68 &    $70^{+5}_{-3}$ &           $19.34\pm0.02$ &          $14.8\pm0.9$ \\
   II &   90427-01-03-02 &   53385.3 &       12272 &  Y &           1.02 &    $70^{+4}_{-5}$ &           $19.04\pm0.02$ &          $14.7\pm0.9$ \\
    - &   90014-01-02-03 &   53385.9 &        1200 &  Y &           1.26 &    $75^{+6}_{-7}$ &           $18.34\pm0.04$ &          $15.0\pm1.3$ \\
  III &  90427-01-03-14G &   53385.9 &       12512 &  Y &           0.70 &    $68^{+4}_{-5}$ &           $18.39\pm0.02$ &          $13.7\pm0.9$ \\
  III &   90014-01-02-00 &   53386.4 &        8240 &  Y &           0.76 &    $71^{+6}_{-4}$ &           $17.36\pm0.02$ &          $13.5\pm0.9$ \\
  III &   90427-01-03-05 &   53386.9 &       12944 &  Y &           0.94 &    $76^{+5}_{-4}$ &           $17.18\pm0.02$ &          $14.3\pm0.8$ \\
    - &   90014-01-02-10 &   53387.1 &        2784 &  Y &           0.92 &    $79^{+7}_{-4}$ &           $16.92\pm0.03$ &          $14.7\pm1.0$ \\
   IV &   90427-01-03-06 &   53387.3 &       10880 &  Y &           1.17 &    $77^{+5}_{-4}$ &           $16.46\pm0.02$ &          $13.9\pm0.8$ \\
   IV &   90427-01-03-07 &   53387.8 &        9632 &  Y &           1.10 &    $76^{+5}_{-4}$ &           $16.07\pm0.02$ &          $13.4\pm0.8$ \\
   IV &   90014-01-02-08 &   53388.0 &        3264 &  Y &           1.31 &          $78\pm6$ &           $15.82\pm0.02$ &          $13.5\pm0.9$ \\
    V &   90427-01-03-09 &   53388.3 &       10912 &  Y &           1.06 &          $81\pm4$ &           $15.48\pm0.02$ &          $13.7\pm0.8$ \\
    V &   90427-01-03-11 &   53388.9 &        9744 &  Y &           0.67 &          $77\pm5$ &           $15.16\pm0.02$ &          $12.8\pm0.8$ \\
    - &   90014-01-02-15 &   53389.1 &        2672 &  Y &           1.11 &    $84^{+6}_{-5}$ &           $15.28\pm0.03$ &          $14.0\pm0.9$ \\
    V &   90427-01-03-12 &   53389.2 &        9664 &  Y &           0.85 &          $85\pm5$ &           $15.07\pm0.02$ &  $14.0^{+0.7}_{-0.8}$ \\
   VI &   90014-01-02-13 &   53390.3 &        7056 &  Y &           1.35 &          $85\pm5$ &           $14.43\pm0.02$ &  $13.3^{+0.7}_{-0.8}$ \\
   VI &   90014-01-03-00 &   53391.3 &        2336 &  Y &           1.28 &    $87^{+7}_{-5}$ &  $13.65^{+0.02}_{-0.01}$ &  $13.0^{+0.8}_{-0.9}$ \\
   VI &   90014-01-03-01 &   53393.2 &        2768 &  Y &           1.08 &    $89^{+6}_{-7}$ &  $12.68^{+0.02}_{-0.01}$ &  $12.2^{+0.6}_{-0.8}$ \\
  VII &  90014-01-03-020 &   53394.3 &       13568 &  Y &           0.98 &          $85\pm6$ &  $12.44^{+0.02}_{-0.01}$ &  $11.5^{+0.3}_{-0.6}$ \\
  VII &   90014-01-03-02 &   53394.6 &        2032 &  Y &           0.98 &    $79^{+7}_{-6}$ &           $12.26\pm0.02$ &          $10.5\pm0.9$ \\
  VII &   90014-01-03-03 &   53395.3 &        6192 &  Y &           1.08 &    $87^{+5}_{-6}$ &           $11.50\pm0.01$ &  $10.8^{+0.5}_{-0.6}$ \\
    - &   90014-01-04-00 &   53398.5 &        1904 &  Y &           1.03 &          $94\pm7$ &            $9.81\pm0.02$ &          $10.0\pm0.8$ \\
    - &   90014-01-04-01 &   53399.6 &         784 &  Y &           0.73 &   $81^{+6}_{-12}$ &            $9.67\pm0.03$ &   $8.5^{+0.9}_{-1.0}$ \\
    - &   90014-01-04-02 &   53401.4 &         944 &  Y &           1.23 &  $82^{+11}_{-10}$ &            $8.90\pm0.02$ &           $7.9\pm0.7$ \\
    - &   90014-01-04-03 &   53403.3 &         624 &  Y &           1.14 &  $78^{+12}_{-14}$ &   $8.08^{+0.03}_{-0.02}$ &           $6.7\pm0.8$ \\
    - &   90014-01-05-00 &   53405.2 &         848 &                       - &           0.83 &   $83^{+10}_{-9}$ &   $8.05^{+0.03}_{-0.02}$ &           $7.1\pm0.7$ \\
    - &   90014-01-05-01 &   53407.6 &        7376 &                       - &           1.87 &    $85^{+6}_{-4}$ &            $7.05\pm0.01$ &           $6.4\pm0.3$ \\
    - &   90014-01-05-04 &   53407.8 &        1472 &                       - &           0.82 &    $79^{+7}_{-8}$ &            $7.24\pm0.02$ &           $6.1\pm0.7$ \\
    - &   90014-01-05-05 &   53408.0 &        1920 &                       - &           1.12 &          $82\pm9$ &   $6.84^{+0.02}_{-0.01}$ &           $5.9\pm0.4$ \\
    - &   90014-01-05-02 &   53409.3 &        2864 &                       - &           2.59 &    $77^{+8}_{-6}$ &            $6.34\pm0.01$ &           $5.1\pm0.4$ \\
    - &   90014-01-05-06 &   53411.6 &        1040 &                       - &           0.76 &   $65^{+12}_{-7}$ &   $5.31^{+0.02}_{-0.01}$ &   $3.6^{+0.4}_{-0.6}$ \\
    - &   90427-01-04-00 &   53413.1 &        5616 &                       - &           1.52 &    $62^{+5}_{-6}$ &            $4.88\pm0.01$ &           $3.2\pm0.2$ \\
    - &   90427-01-04-04 &   53413.7 &        6016 &                       - &           1.62 &    $61^{+6}_{-5}$ &            $4.56\pm0.01$ &           $2.9\pm0.2$ \\
    - &   90014-01-06-00 &   53414.0 &        1200 &                       - &           1.04 &    $54^{+8}_{-7}$ &            $4.43\pm0.01$ &           $2.5\pm0.3$ \\
    - &   90427-01-04-02 &   53414.2 &       11392 &                       - &           1.67 &          $62\pm5$ &            $4.24\pm0.01$ &           $2.7\pm0.2$ \\
    - &   90427-01-04-03 &   53414.5 &        6736 &                       - &           1.46 &    $65^{+5}_{-4}$ &            $4.02\pm0.00$ &           $2.7\pm0.2$ \\
    - &   90427-01-04-05 &   53414.8 &        2128 &                       - &           1.94 &   $55^{+8}_{-10}$ &            $3.89\pm0.01$ &   $2.2^{+0.2}_{-0.3}$ \\
    - &   90014-01-06-01 &   53416.1 &        1872 &                       - &           1.06 &  $67^{+13}_{-10}$ &            $3.37\pm0.01$ &           $2.3\pm0.4$ \\
    - &   90427-01-04-01 &   53416.5 &        5264 &                       - &           0.97 &    $56^{+7}_{-6}$ &   $3.32^{+0.00}_{-0.01}$ &           $1.9\pm0.2$ \\
    - &   90014-01-06-02 &   53417.6 &        1248 &                       - &           1.21 &  $64^{+12}_{-14}$ &            $2.91\pm0.01$ &           $1.9\pm0.3$ \\
    - &   90014-01-06-03 &   53418.5 &        1616 &                       - &           0.73 &  $39^{+12}_{-14}$ &            $2.49\pm0.01$ &           $1.0\pm0.3$ \\
    - &   90014-01-07-01 &   53419.4 &        1568 &                       - &           1.56 &  $47^{+13}_{-11}$ &            $2.16\pm0.01$ &           $1.1\pm0.3$ \\
    - &   90014-01-07-03 &   53420.7 &        1504 &                       - &           1.48 &         $42\pm14$ &            $1.90\pm0.01$ &           $0.8\pm0.3$ \\
 VIII &   90014-01-07-04 &   53422.6 &        1888 &                       Z &           1.20 &  $<27$ &            $1.37\pm0.01$ &   $<0.4$ \\
 VIII &   90014-01-07-00 &   53424.4 &        2864 &                       Z &           0.61 &    $<38$ &            $0.93\pm0.00$ &   $<0.1$ \\
    - &   90014-01-08-00 &   53426.5 &        2384 &                       - &           0.76 &    $<87$ &            $0.55\pm0.01$ &   $<0.2$ \\
    - &   90014-01-08-01 &   53428.5 &        2160 &                       - &           1.19 &  $<60$ &            $0.37\pm0.01$ &   $<0.2$ \\
    - &   90014-01-08-02 &   53430.5 &        1968 &                       - &           0.69 &    $<16$ &            $0.01\pm0.00$ &   $<0.1$ \\
    - &   90014-01-08-03 &   53432.4 &        2176 &                       - &           0.32 &    $<83$ &            $0.00\pm0.00$ &   $<0.1$ 
\end{longtable}
*: X is for B\_16ms\_64M\_0\_249\_H configuration, Y is for B\_16ms\_46M\_0\_49\_H, Z is for E\_125us\_64M\_0\_1s
  \twocolumn

%% file: pulsar.bbl
\begin{thebibliography}{}
\makeatletter
\relax
\def\mn@urlcharsother{\let\do\@makeother \do\$\do\&\do\#\do\^\do\_\do\%\do\~}
\def\mn@doi{\begingroup\mn@urlcharsother \@ifnextchar [ {\mn@doi@}
  {\mn@doi@[]}}
\def\mn@doi@[#1]#2{\def\@tempa{#1}\ifx\@tempa\@empty \href
  {http://dx.doi.org/#2} {doi:#2}\else \href {http://dx.doi.org/#2} {#1}\fi
  \endgroup}
\def\mn@eprint#1#2{\mn@eprint@#1:#2::\@nil}
\def\mn@eprint@arXiv#1{\href {http://arxiv.org/abs/#1} {{\tt arXiv:#1}}}
\def\mn@eprint@dblp#1{\href {http://dblp.uni-trier.de/rec/bibtex/#1.xml}
  {dblp:#1}}
\def\mn@eprint@#1:#2:#3:#4\@nil{\def\@tempa {#1}\def\@tempb {#2}\def\@tempc
  {#3}\ifx \@tempc \@empty \let \@tempc \@tempb \let \@tempb \@tempa \fi \ifx
  \@tempb \@empty \def\@tempb {arXiv}\fi \@ifundefined
  {mn@eprint@\@tempb}{\@tempb:\@tempc}{\expandafter \expandafter \csname
  mn@eprint@\@tempb\endcsname \expandafter{\@tempc}}}

\bibitem[\protect\citeauthoryear{{Aftab}, {Paul}  \& {Kretschmar}}{{Aftab}
  et~al.}{2019}]{Aftab2019}
{Aftab} N.,  {Paul} B.,   {Kretschmar} P.,  2019, \mn@doi [\apjs]
  {10.3847/1538-4365/ab2a77}, \href
  {https://ui.adsabs.harvard.edu/abs/2019ApJS..243...29A} {243, 29}

\bibitem[\protect\citeauthoryear{{Arnason}, {Papei}, {Barmby}, {Bahramian}  \&
  {Gorski}}{{Arnason} et~al.}{2021}]{Arnason2021}
{Arnason} R.~M.,  {Papei} H.,  {Barmby} P.,  {Bahramian} A.,   {Gorski} M.~D.,
  2021, arXiv e-prints, \href
  {https://ui.adsabs.harvard.edu/abs/2021arXiv210202615A} {p. arXiv:2102.02615}

\bibitem[\protect\citeauthoryear{{Arnaud}}{{Arnaud}}{1996}]{Arnaud1996}
{Arnaud} K.~A.,  1996, {XSPEC: The First Ten Years}.
p.~17

\bibitem[\protect\citeauthoryear{{Basko} \& {Sunyaev}}{{Basko} \&
  {Sunyaev}}{1976}]{Basko1976}
{Basko} M.~M.,  {Sunyaev} R.~A.,  1976, \mn@doi [\mnras]
  {10.1093/mnras/175.2.395}, \href
  {https://ui.adsabs.harvard.edu/abs/1976MNRAS.175..395B} {175, 395}

\bibitem[\protect\citeauthoryear{{Basko}, {Sunyaev}  \& {Titarchuk}}{{Basko}
  et~al.}{1974}]{Basko1974}
{Basko} M.~M.,  {Sunyaev} R.~A.,   {Titarchuk} L.~G.,  1974, \aap, \href
  {https://ui.adsabs.harvard.edu/abs/1974A&A....31..249B} {31, 249}

\bibitem[\protect\citeauthoryear{{Baum}, {Cherry}  \& {Rodi}}{{Baum}
  et~al.}{2017}]{Baum2017}
{Baum} Z.~A.,  {Cherry} M.~L.,   {Rodi} J.,  2017, \mn@doi [\mnras]
  {10.1093/mnras/stx384}, \href
  {https://ui.adsabs.harvard.edu/abs/2017MNRAS.467.4424B} {467, 4424}

\bibitem[\protect\citeauthoryear{{Caballero-Garc{\'\i}a}
  et~al.,}{{Caballero-Garc{\'\i}a} et~al.}{2016}]{Caballero2016}
{Caballero-Garc{\'\i}a} M.~D.,  et~al., 2016, \mn@doi [\aap]
  {10.1051/0004-6361/201526849}, \href
  {https://ui.adsabs.harvard.edu/abs/2016A&A...589A...9C} {589, A9}

\bibitem[\protect\citeauthoryear{{Choi}, {Nagase}, {Makino}, {Dotani},
  {Kitamoto}  \& {Takahama}}{{Choi} et~al.}{1994}]{Choi1994}
{Choi} C.~S.,  {Nagase} F.,  {Makino} F.,  {Dotani} T.,  {Kitamoto} S.,
  {Takahama} S.,  1994, \mn@doi [\apj] {10.1086/175008}, \href
  {https://ui.adsabs.harvard.edu/abs/1994ApJ...437..449C} {437, 449}

\bibitem[\protect\citeauthoryear{{Churazov}, {Sunyaev}, {Gilfanov}, {Forman}
  \& {Jones}}{{Churazov} et~al.}{1998}]{Churazov1998}
{Churazov} E.,  {Sunyaev} R.,  {Gilfanov} M.,  {Forman} W.,   {Jones} C.,
  1998, \mn@doi [\mnras] {10.1046/j.1365-8711.1998.01685.x}, \href
  {https://ui.adsabs.harvard.edu/abs/1998MNRAS.297.1274C} {297, 1274}

\bibitem[\protect\citeauthoryear{{Churazov}, {Gilfanov}  \&
  {Revnivtsev}}{{Churazov} et~al.}{2001}]{Churazov2001}
{Churazov} E.,  {Gilfanov} M.,   {Revnivtsev} M.,  2001, \mn@doi [\mnras]
  {10.1046/j.1365-8711.2001.04056.x}, \href
  {https://ui.adsabs.harvard.edu/abs/2001MNRAS.321..759C} {321, 759}

\bibitem[\protect\citeauthoryear{{Coe} \& {Kirk}}{{Coe} \&
  {Kirk}}{2015}]{Coe2015}
{Coe} M.~J.,  {Kirk} J.,  2015, \mn@doi [\mnras] {10.1093/mnras/stv1283}, \href
  {https://ui.adsabs.harvard.edu/abs/2015MNRAS.452..969C} {452, 969}

\bibitem[\protect\citeauthoryear{{Cusumano}, {La Parola}, {D'A{\`\i}},
  {Segreto}, {Tagliaferri}, {Barthelmy}  \& {Gehrels}}{{Cusumano}
  et~al.}{2016}]{Cusumano2016}
{Cusumano} G.,  {La Parola} V.,  {D'A{\`\i}} A.,  {Segreto} A.,  {Tagliaferri}
  G.,  {Barthelmy} S.~D.,   {Gehrels} N.,  2016, \mn@doi [\mnras]
  {10.1093/mnrasl/slw084}, \href
  {https://ui.adsabs.harvard.edu/abs/2016MNRAS.460L..99C} {460, L99}

\bibitem[\protect\citeauthoryear{{Day}, {Nagase}, {Asai}  \& {Takeshima}}{{Day}
  et~al.}{1993}]{Day1993}
{Day} C.~S.~R.,  {Nagase} F.,  {Asai} K.,   {Takeshima} T.,  1993, \mn@doi
  [\apj] {10.1086/172625}, \href
  {https://ui.adsabs.harvard.edu/abs/1993ApJ...408..656D} {408, 656}

\bibitem[\protect\citeauthoryear{{Doroshenko}, {Tsygankov}  \&
  {Santangelo}}{{Doroshenko} et~al.}{2016}]{Doroshenko2016}
{Doroshenko} V.,  {Tsygankov} S.,   {Santangelo} A.,  2016, \mn@doi [\aap]
  {10.1051/0004-6361/201527756}, \href
  {https://ui.adsabs.harvard.edu/abs/2016A&A...589A..72D} {589, A72}

\bibitem[\protect\citeauthoryear{{Doroshenko}, {Tsygankov}, {Mushtukov},
  {Lutovinov}, {Santangelo}, {Suleimanov}  \& {Poutanen}}{{Doroshenko}
  et~al.}{2017}]{Doroshenko2017}
{Doroshenko} V.,  {Tsygankov} S.~S.,  {Mushtukov} A. e.~A.,  {Lutovinov} A.~A.,
   {Santangelo} A.,  {Suleimanov} V.~F.,   {Poutanen} J.,  2017, \mn@doi
  [\mnras] {10.1093/mnras/stw3236}, \href
  {https://ui.adsabs.harvard.edu/abs/2017MNRAS.466.2143D} {466, 2143}

\bibitem[\protect\citeauthoryear{{Endo}, {Ishida}, {Masai}, {Kunieda}, {Inoue}
  \& {Nagase}}{{Endo} et~al.}{2002}]{Endo2002}
{Endo} T.,  {Ishida} M.,  {Masai} K.,  {Kunieda} H.,  {Inoue} H.,   {Nagase}
  F.,  2002, \mn@doi [\apj] {10.1086/341060}, \href
  {https://ui.adsabs.harvard.edu/abs/2002ApJ...574..879E} {574, 879}

\bibitem[\protect\citeauthoryear{{Epili}, {Naik}, {Jaisawal}  \&
  {Gupta}}{{Epili} et~al.}{2017}]{Epili2017}
{Epili} P.,  {Naik} S.,  {Jaisawal} G.~K.,   {Gupta} S.,  2017, \mn@doi
  [\mnras] {10.1093/mnras/stx2247}, \href
  {https://ui.adsabs.harvard.edu/abs/2017MNRAS.472.3455E} {472, 3455}

\bibitem[\protect\citeauthoryear{{Fabian}, {Rees}, {Stella}  \&
  {White}}{{Fabian} et~al.}{1989}]{Fabian1989}
{Fabian} A.~C.,  {Rees} M.~J.,  {Stella} L.,   {White} N.~E.,  1989, \mn@doi
  [\mnras] {10.1093/mnras/238.3.729}, \href
  {https://ui.adsabs.harvard.edu/abs/1989MNRAS.238..729F} {238, 729}

\bibitem[\protect\citeauthoryear{{Filippova}, {Mereminskiy}, {Lutovinov},
  {Molkov}  \& {Tsygankov}}{{Filippova} et~al.}{2017}]{Filippova2017}
{Filippova} E.~V.,  {Mereminskiy} I.~A.,  {Lutovinov} A.~A.,  {Molkov} S.~V.,
  {Tsygankov} S.~S.,  2017, \mn@doi [Astronomy Letters]
  {10.1134/S1063773717110020}, \href
  {https://ui.adsabs.harvard.edu/abs/2017AstL...43..706F} {43, 706}

\bibitem[\protect\citeauthoryear{{Frank}, {King}  \& {Raine}}{{Frank}
  et~al.}{2002}]{Frank2002}
{Frank} J.,  {King} A.,   {Raine} D.~J.,  2002, {Accretion Power in
  Astrophysics: Third Edition}

\bibitem[\protect\citeauthoryear{{Garc{\'\i}a}, {McClintock}, {Steiner},
  {Remillard}  \& {Grinberg}}{{Garc{\'\i}a} et~al.}{2014}]{Garcia2014}
{Garc{\'\i}a} J.~A.,  {McClintock} J.~E.,  {Steiner} J.~F.,  {Remillard} R.~A.,
    {Grinberg} V.,  2014, \mn@doi [\apj] {10.1088/0004-637X/794/1/73}, \href
  {https://ui.adsabs.harvard.edu/abs/2014ApJ...794...73G} {794, 73}

\bibitem[\protect\citeauthoryear{{George} \& {Fabian}}{{George} \&
  {Fabian}}{1991}]{George1991}
{George} I.~M.,  {Fabian} A.~C.,  1991, \mn@doi [\mnras]
  {10.1093/mnras/249.2.352}, \href
  {https://ui.adsabs.harvard.edu/abs/1991MNRAS.249..352G} {249, 352}

\bibitem[\protect\citeauthoryear{{Gilfanov}}{{Gilfanov}}{2010}]{Gilfanov2010}
{Gilfanov} M.,  2010, {X-Ray Emission from Black-Hole Binaries}.
p.~17, \mn@doi{10.1007/978-3-540-76937-8_2}

\bibitem[\protect\citeauthoryear{{Gilfanov}, {Churazov}  \&
  {Revnivtsev}}{{Gilfanov} et~al.}{1999}]{Gilfanov1999}
{Gilfanov} M.,  {Churazov} E.,   {Revnivtsev} M.,  1999, \aap, \href
  {https://ui.adsabs.harvard.edu/abs/1999A&A...352..182G} {352, 182}

\bibitem[\protect\citeauthoryear{{Gilfanov}, {Churazov}  \&
  {Revnivtsev}}{{Gilfanov} et~al.}{2000}]{Gilfanov2000}
{Gilfanov} M.,  {Churazov} E.,   {Revnivtsev} M.,  2000, \mn@doi [\mnras]
  {10.1046/j.1365-8711.2000.03686.x}, \href
  {https://ui.adsabs.harvard.edu/abs/2000MNRAS.316..923G} {316, 923}

\bibitem[\protect\citeauthoryear{{Gim{\'e}nez-Garc{\'\i}a}, {Torrej{\'o}n},
  {Eikmann}, {Mart{\'\i}nez-N{\'u}{\~n}ez}, {Oskinova}, {Rodes-Roca}  \&
  {Bernab{\'e}u}}{{Gim{\'e}nez-Garc{\'\i}a} et~al.}{2015}]{Gimenez-Garcia2015}
{Gim{\'e}nez-Garc{\'\i}a} A.,  {Torrej{\'o}n} J.~M.,  {Eikmann} W.,
  {Mart{\'\i}nez-N{\'u}{\~n}ez} S.,  {Oskinova} L.~M.,  {Rodes-Roca} J.~J.,
  {Bernab{\'e}u} G.,  2015, \mn@doi [\aap] {10.1051/0004-6361/201425004}, \href
  {https://ui.adsabs.harvard.edu/abs/2015A&A...576A.108G} {576, A108}

\bibitem[\protect\citeauthoryear{{Gnedin} \& {Sunyaev}}{{Gnedin} \&
  {Sunyaev}}{1973}]{Gnedin1973}
{Gnedin} Y.~N.,  {Sunyaev} R.~A.,  1973, \aap, \href
  {https://ui.adsabs.harvard.edu/abs/1973A&A....25..233G} {25, 233}

\bibitem[\protect\citeauthoryear{{Gupta}, {Naik}, {Jaisawal}  \&
  {Epili}}{{Gupta} et~al.}{2018}]{Gupta2018}
{Gupta} S.,  {Naik} S.,  {Jaisawal} G.~K.,   {Epili} P.~R.,  2018, \mn@doi
  [\mnras] {10.1093/mnras/sty1804}, \href
  {https://ui.adsabs.harvard.edu/abs/2018MNRAS.479.5612G} {479, 5612}

\bibitem[\protect\citeauthoryear{{Hemphill}, {Rothschild}, {Markowitz},
  {F{\"u}rst}, {Pottschmidt}  \& {Wilms}}{{Hemphill}
  et~al.}{2014}]{Hemphill2014}
{Hemphill} P.~B.,  {Rothschild} R.~E.,  {Markowitz} A.,  {F{\"u}rst} F.,
  {Pottschmidt} K.,   {Wilms} J.,  2014, \mn@doi [\apj]
  {10.1088/0004-637X/792/1/14}, \href
  {https://ui.adsabs.harvard.edu/abs/2014ApJ...792...14H} {792, 14}

\bibitem[\protect\citeauthoryear{{Honeycutt} \& {Schlegel}}{{Honeycutt} \&
  {Schlegel}}{1985}]{Honeycutt1985}
{Honeycutt} R.~K.,  {Schlegel} E.~M.,  1985, \mn@doi [\pasp] {10.1086/131534},
  \href {https://ui.adsabs.harvard.edu/abs/1985PASP...97..300H} {97, 300}

\bibitem[\protect\citeauthoryear{{Inoue}}{{Inoue}}{1985}]{Inoue1985}
{Inoue} H.,  1985, \mn@doi [\ssr] {10.1007/BF00212905}, \href
  {https://ui.adsabs.harvard.edu/abs/1985SSRv...40..317I} {40, 317}

\bibitem[\protect\citeauthoryear{{Jahoda}, {Swank}, {Giles}, {Stark},
  {Strohmayer}, {Zhang}  \& {Morgan}}{{Jahoda} et~al.}{1996}]{Jahoda1996}
{Jahoda} K.,  {Swank} J.~H.,  {Giles} A.~B.,  {Stark} M.~J.,  {Strohmayer} T.,
  {Zhang} W.,   {Morgan} E.~H.,  1996, {In-orbit performance and calibration of
  the Rossi X-ray Timing Explorer (RXTE) Proportional Counter Array (PCA)}.
pp 59--70, \mn@doi{10.1117/12.256034}

\bibitem[\protect\citeauthoryear{{Jaisawal} \& {Naik}}{{Jaisawal} \&
  {Naik}}{2014}]{Jaisawal2014}
{Jaisawal} G.~K.,  {Naik} S.,  2014, Bulletin of the Astronomical Society of
  India, \href {https://ui.adsabs.harvard.edu/abs/2014BASI...42..147J} {42,
  147}

\bibitem[\protect\citeauthoryear{{Jaisawal}, {Naik}  \& {Epili}}{{Jaisawal}
  et~al.}{2016}]{Jaisawal2016}
{Jaisawal} G.~K.,  {Naik} S.,   {Epili} P.,  2016, \mn@doi [\mnras]
  {10.1093/mnras/stw085}, \href
  {https://ui.adsabs.harvard.edu/abs/2016MNRAS.457.2749J} {457, 2749}

\bibitem[\protect\citeauthoryear{{Jaisawal}, {Naik}, {Ho}, {Kumari}, {Epili}
  \& {Vasilopoulos}}{{Jaisawal} et~al.}{2020}]{Jaisawal2020}
{Jaisawal} G.~K.,  {Naik} S.,  {Ho} W. C.~G.,  {Kumari} N.,  {Epili} P.,
  {Vasilopoulos} G.,  2020, \mn@doi [\mnras] {10.1093/mnras/staa2604}, \href
  {https://ui.adsabs.harvard.edu/abs/2020MNRAS.498.4830J} {498, 4830}

\bibitem[\protect\citeauthoryear{{Jaisawal}, {Naik}, {Epili}, {Chhotaray},
  {Jana}  \& {Agrawal}}{{Jaisawal} et~al.}{2021}]{Jaisawal2021}
{Jaisawal} G.~K.,  {Naik} S.,  {Epili} P.~R.,  {Chhotaray} B.,  {Jana} A.,
  {Agrawal} P.~C.,  2021, arXiv e-prints, \href
  {https://ui.adsabs.harvard.edu/abs/2021arXiv210100815J} {p. arXiv:2101.00815}

\bibitem[\protect\citeauthoryear{{Ji} et~al.,}{{Ji} et~al.}{2021}]{Ji2021}
{Ji} L.,  et~al., 2021, \mn@doi [\mnras] {10.1093/mnras/staa3788}, \href
  {https://ui.adsabs.harvard.edu/abs/2021MNRAS.501.2522J} {501, 2522}

\bibitem[\protect\citeauthoryear{{Kohmura}, {Kitamoto}  \& {Torii}}{{Kohmura}
  et~al.}{2001}]{Kohmura2001}
{Kohmura} T.,  {Kitamoto} S.,   {Torii} K.,  2001, \mn@doi [\apj]
  {10.1086/323848}, \href
  {https://ui.adsabs.harvard.edu/abs/2001ApJ...562..943K} {562, 943}

\bibitem[\protect\citeauthoryear{{Koliopanos} \& {Gilfanov}}{{Koliopanos} \&
  {Gilfanov}}{2016}]{Koliopanos2016}
{Koliopanos} F.,  {Gilfanov} M.,  2016, \mn@doi [\mnras]
  {10.1093/mnras/stv2873}, \href
  {https://ui.adsabs.harvard.edu/abs/2016MNRAS.456.3535K} {456, 3535}

\bibitem[\protect\citeauthoryear{{Kreykenbohm} et~al.,}{{Kreykenbohm}
  et~al.}{2005}]{Kreykenbohm2005}
{Kreykenbohm} I.,  et~al., 2005, \mn@doi [\aap] {10.1051/0004-6361:200500023},
  \href {https://ui.adsabs.harvard.edu/abs/2005A&A...433L..45K} {433, L45}

\bibitem[\protect\citeauthoryear{{Leahy}, {Matsuoka}, {Kawai}  \&
  {Makino}}{{Leahy} et~al.}{1989}]{Leahy1989}
{Leahy} D.~A.,  {Matsuoka} M.,  {Kawai} N.,   {Makino} F.,  1989, \mn@doi
  [\mnras] {10.1093/mnras/236.3.603}, \href
  {https://ui.adsabs.harvard.edu/abs/1989MNRAS.236..603L} {236, 603}

\bibitem[\protect\citeauthoryear{{Liu}, {Soria}, {Qiao}  \& {Liu}}{{Liu}
  et~al.}{2018}]{Liu2018}
{Liu} J.,  {Soria} R.,  {Qiao} E.,   {Liu} J.,  2018, \mn@doi [\mnras]
  {10.1093/mnras/sty2180}, \href
  {https://ui.adsabs.harvard.edu/abs/2018MNRAS.480.4746L} {480, 4746}

\bibitem[\protect\citeauthoryear{{Liu} et~al.,}{{Liu} et~al.}{2021}]{Liu2021}
{Liu} J.,  et~al., 2021, \mn@doi [\mnras] {10.1093/mnras/stab938}, \href
  {https://ui.adsabs.harvard.edu/abs/2021MNRAS.tmp..917L} {}

\bibitem[\protect\citeauthoryear{{Lutovinov}, {Tsygankov}, {Suleimanov},
  {Mushtukov}, {Doroshenko}, {Nagirner}  \& {Poutanen}}{{Lutovinov}
  et~al.}{2015}]{Lutovinov2015}
{Lutovinov} A.~A.,  {Tsygankov} S.~S.,  {Suleimanov} V.~F.,  {Mushtukov} A.~A.,
   {Doroshenko} V.,  {Nagirner} D.~I.,   {Poutanen} J.,  2015, \mn@doi [\mnras]
  {10.1093/mnras/stv125}, \href
  {https://ui.adsabs.harvard.edu/abs/2015MNRAS.448.2175L} {448, 2175}

\bibitem[\protect\citeauthoryear{{Lutovinov} et~al.,}{{Lutovinov}
  et~al.}{2021}]{Lutovinov2021}
{Lutovinov} A.,  et~al., 2021, arXiv e-prints, \href
  {https://ui.adsabs.harvard.edu/abs/2021arXiv210305728L} {p. arXiv:2103.05728}

\bibitem[\protect\citeauthoryear{{Makishima}}{{Makishima}}{1986}]{Makishima1986}
{Makishima} K.,  1986, {Iron Lines from Galactic and Extragalactic X-ray
  Sources}.
p.~249, \mn@doi{10.1007/3-540-17195-9_14}

\bibitem[\protect\citeauthoryear{{Makishima} et~al.,}{{Makishima}
  et~al.}{1990}]{Makishima1990}
{Makishima} K.,  et~al., 1990, \mn@doi [\apjl] {10.1086/185888}, \href
  {https://ui.adsabs.harvard.edu/abs/1990ApJ...365L..59M} {365, L59}

\bibitem[\protect\citeauthoryear{{Mushtukov}, {Suleimanov}, {Tsygankov}  \&
  {Poutanen}}{{Mushtukov} et~al.}{2015a}]{Mushtukov2015b}
{Mushtukov} A.~A.,  {Suleimanov} V.~F.,  {Tsygankov} S.~S.,   {Poutanen} J.,
  2015a, \mn@doi [\mnras] {10.1093/mnras/stu2484}, \href
  {https://ui.adsabs.harvard.edu/abs/2015MNRAS.447.1847M} {447, 1847}

\bibitem[\protect\citeauthoryear{{Mushtukov}, {Suleimanov}, {Tsygankov}  \&
  {Poutanen}}{{Mushtukov} et~al.}{2015b}]{Mushtukov2015}
{Mushtukov} A.~A.,  {Suleimanov} V.~F.,  {Tsygankov} S.~S.,   {Poutanen} J.,
  2015b, \mn@doi [\mnras] {10.1093/mnras/stv2087}, \href
  {https://ui.adsabs.harvard.edu/abs/2015MNRAS.454.2539M} {454, 2539}

\bibitem[\protect\citeauthoryear{{Mushtukov}, {Suleimanov}, {Tsygankov}  \&
  {Ingram}}{{Mushtukov} et~al.}{2017}]{Mushtukov2017}
{Mushtukov} A.~A.,  {Suleimanov} V.~F.,  {Tsygankov} S.~S.,   {Ingram} A.,
  2017, \mn@doi [\mnras] {10.1093/mnras/stx141}, \href
  {https://ui.adsabs.harvard.edu/abs/2017MNRAS.467.1202M} {467, 1202}

\bibitem[\protect\citeauthoryear{{Mushtukov}, {Verhagen}, {Tsygankov}, {van der
  Klis}, {Lutovinov}  \& {Larchenkova}}{{Mushtukov}
  et~al.}{2018}]{Mushtukov2018}
{Mushtukov} A.~A.,  {Verhagen} P.~A.,  {Tsygankov} S.~S.,  {van der Klis} M.,
  {Lutovinov} A.~A.,   {Larchenkova} T.~I.,  2018, \mn@doi [\mnras]
  {10.1093/mnras/stx2905}, \href
  {https://ui.adsabs.harvard.edu/abs/2018MNRAS.474.5425M} {474, 5425}

\bibitem[\protect\citeauthoryear{{Nagase}, {Corbet}, {Day}, {Inoue},
  {Takeshima}, {Yoshida}  \& {Mihara}}{{Nagase} et~al.}{1992}]{Nagase1992}
{Nagase} F.,  {Corbet} R.~H.~D.,  {Day} C.~S.~R.,  {Inoue} H.,  {Takeshima} T.,
   {Yoshida} K.,   {Mihara} T.,  1992, \mn@doi [\apj] {10.1086/171705}, \href
  {https://ui.adsabs.harvard.edu/abs/1992ApJ...396..147N} {396, 147}

\bibitem[\protect\citeauthoryear{{Naik} et~al.,}{{Naik}
  et~al.}{2008}]{Naik2008}
{Naik} S.,  et~al., 2008, \mn@doi [\apj] {10.1086/523295}, \href
  {https://ui.adsabs.harvard.edu/abs/2008ApJ...672..516N} {672, 516}

\bibitem[\protect\citeauthoryear{{Naik}, {Paul}, {Kachhara}  \&
  {Vadawale}}{{Naik} et~al.}{2011}]{Naik2011}
{Naik} S.,  {Paul} B.,  {Kachhara} C.,   {Vadawale} S.~V.,  2011, \mn@doi
  [\mnras] {10.1111/j.1365-2966.2010.18128.x}, \href
  {https://ui.adsabs.harvard.edu/abs/2011MNRAS.413..241N} {413, 241}

\bibitem[\protect\citeauthoryear{{Negueruela}, {Roche}, {Fabregat}  \&
  {Coe}}{{Negueruela} et~al.}{1999}]{Negueruela1999}
{Negueruela} I.,  {Roche} P.,  {Fabregat} J.,   {Coe} M.~J.,  1999, \mn@doi
  [\mnras] {10.1046/j.1365-8711.1999.02682.x}, \href
  {https://ui.adsabs.harvard.edu/abs/1999MNRAS.307..695N} {307, 695}

\bibitem[\protect\citeauthoryear{{Nespoli} \& {Reig}}{{Nespoli} \&
  {Reig}}{2011}]{Nespoli2011}
{Nespoli} E.,  {Reig} P.,  2011, \mn@doi [\aap] {10.1051/0004-6361/201015303},
  \href {https://ui.adsabs.harvard.edu/abs/2011A&A...526A...7N} {526, A7}

\bibitem[\protect\citeauthoryear{{Okazaki} \& {Negueruela}}{{Okazaki} \&
  {Negueruela}}{2001}]{Okazaki2001}
{Okazaki} A.~T.,  {Negueruela} I.,  2001, {Origin of the X-ray Activity of
  Be/X-ray Binaries}.
p.~281

\bibitem[\protect\citeauthoryear{{Okazaki}, {Bate}, {Ogilvie}  \&
  {Pringle}}{{Okazaki} et~al.}{2002}]{Okazaki2002}
{Okazaki} A.~T.,  {Bate} M.~R.,  {Ogilvie} G.~I.,   {Pringle} J.~E.,  2002,
  \mn@doi [\mnras] {10.1046/j.1365-8711.2002.05960.x}, \href
  {https://ui.adsabs.harvard.edu/abs/2002MNRAS.337..967O} {337, 967}

\bibitem[\protect\citeauthoryear{{Poutanen}, {Mushtukov}, {Suleimanov},
  {Tsygankov}, {Nagirner}, {Doroshenko}  \& {Lutovinov}}{{Poutanen}
  et~al.}{2013}]{Poutanen2013}
{Poutanen} J.,  {Mushtukov} A.~A.,  {Suleimanov} V.~F.,  {Tsygankov} S.~S.,
  {Nagirner} D.~I.,  {Doroshenko} V.,   {Lutovinov} A. e.~A.,  2013, \mn@doi
  [\apj] {10.1088/0004-637X/777/2/115}, \href
  {https://ui.adsabs.harvard.edu/abs/2013ApJ...777..115P} {777, 115}

\bibitem[\protect\citeauthoryear{{Reig}}{{Reig}}{2011}]{Reig2011}
{Reig} P.,  2011, \mn@doi [\apss] {10.1007/s10509-010-0575-8}, \href
  {https://ui.adsabs.harvard.edu/abs/2011Ap&SS.332....1R} {332, 1}

\bibitem[\protect\citeauthoryear{{Rouco Escorial}, {Wijnands}, {Ootes},
  {Degenaar}, {Snelders}, {Kaper}, {Cackett}  \& {Homan}}{{Rouco Escorial}
  et~al.}{2019}]{RoucoEscorial2019}
{Rouco Escorial} A.,  {Wijnands} R.,  {Ootes} L.~S.,  {Degenaar} N.,
  {Snelders} M.,  {Kaper} L.,  {Cackett} E.~M.,   {Homan} J.,  2019, \mn@doi
  [\aap] {10.1051/0004-6361/201834327}, \href
  {https://ui.adsabs.harvard.edu/abs/2019A&A...630A.105R} {630, A105}

\bibitem[\protect\citeauthoryear{{Sanjurjo-Ferr{\'\i}n}, {Torrej{\'o}n},
  {Postnov}, {Oskinova}, {Rodes-Roca}  \& {Bernabeu}}{{Sanjurjo-Ferr{\'\i}n}
  et~al.}{2021}]{Sanjurjo-Ferin2021}
{Sanjurjo-Ferr{\'\i}n} G.,  {Torrej{\'o}n} J.~M.,  {Postnov} K.,  {Oskinova}
  L.,  {Rodes-Roca} J.~J.,   {Bernabeu} G.,  2021, \mn@doi [\mnras]
  {10.1093/mnras/staa3953}, \href
  {https://ui.adsabs.harvard.edu/abs/2021MNRAS.501.5892S} {501, 5892}

\bibitem[\protect\citeauthoryear{{Shtykovsky}, {Lutovinov}, {Arefiev},
  {Molkov}, {Tsygankov}  \& {Revnivtsev}}{{Shtykovsky}
  et~al.}{2017}]{Shtykovsky2017}
{Shtykovsky} A.~E.,  {Lutovinov} A.~A.,  {Arefiev} V.~A.,  {Molkov} S.~V.,
  {Tsygankov} S.~S.,   {Revnivtsev} M.~G.,  2017, \mn@doi [Astronomy Letters]
  {10.1134/S1063773717030069}, \href
  {https://ui.adsabs.harvard.edu/abs/2017AstL...43..175S} {43, 175}

\bibitem[\protect\citeauthoryear{{Stella}, {White}, {Davelaar}, {Parmar},
  {Blissett}  \& {van der Klis}}{{Stella} et~al.}{1985}]{Stella1985}
{Stella} L.,  {White} N.~E.,  {Davelaar} J.,  {Parmar} A.~N.,  {Blissett}
  R.~J.,   {van der Klis} M.,  1985, \mn@doi [\apjl] {10.1086/184419}, \href
  {https://ui.adsabs.harvard.edu/abs/1985ApJ...288L..45S} {288, L45}

\bibitem[\protect\citeauthoryear{{Terrell} \& {Priedhorsky}}{{Terrell} \&
  {Priedhorsky}}{1984}]{Terrell1984}
{Terrell} J.,  {Priedhorsky} W.~C.,  1984, \mn@doi [\apjl] {10.1086/184355},
  \href {https://ui.adsabs.harvard.edu/abs/1984ApJ...285L..15T} {285, L15}

\bibitem[\protect\citeauthoryear{{Tsygankov} \& {Lutovinov}}{{Tsygankov} \&
  {Lutovinov}}{2010}]{Tsygankov2010b}
{Tsygankov} S.,  {Lutovinov} A.,  2010, arXiv e-prints, \href
  {https://ui.adsabs.harvard.edu/abs/2010arXiv1002.1898T} {p. arXiv:1002.1898}

\bibitem[\protect\citeauthoryear{{Tsygankov}, {Lutovinov}, {Churazov}  \&
  {Sunyaev}}{{Tsygankov} et~al.}{2006}]{Tsygankov2006}
{Tsygankov} S.~S.,  {Lutovinov} A.~A.,  {Churazov} E.~M.,   {Sunyaev} R.~A.,
  2006, \mn@doi [\mnras] {10.1111/j.1365-2966.2006.10610.x}, \href
  {https://ui.adsabs.harvard.edu/abs/2006MNRAS.371...19T} {371, 19}

\bibitem[\protect\citeauthoryear{{Tsygankov}, {Lutovinov}, {Churazov}  \&
  {Sunyaev}}{{Tsygankov} et~al.}{2007}]{Tsygankov2007}
{Tsygankov} S.~S.,  {Lutovinov} A.~A.,  {Churazov} E.~M.,   {Sunyaev} R.~A.,
  2007, \mn@doi [Astronomy Letters] {10.1134/S1063773707060023}, \href
  {https://ui.adsabs.harvard.edu/abs/2007AstL...33..368T} {33, 368}

\bibitem[\protect\citeauthoryear{{Tsygankov}, {Lutovinov}  \&
  {Serber}}{{Tsygankov} et~al.}{2010}]{Tsygankov2010}
{Tsygankov} S.~S.,  {Lutovinov} A.~A.,   {Serber} A.~V.,  2010, \mn@doi
  [\mnras] {10.1111/j.1365-2966.2009.15791.x}, \href
  {https://ui.adsabs.harvard.edu/abs/2010MNRAS.401.1628T} {401, 1628}

\bibitem[\protect\citeauthoryear{{Tsygankov}, {Lutovinov}, {Doroshenko},
  {Mushtukov}, {Suleimanov}  \& {Poutanen}}{{Tsygankov}
  et~al.}{2016}]{Tsygankov2016}
{Tsygankov} S.~S.,  {Lutovinov} A.~A.,  {Doroshenko} V.,  {Mushtukov} A.~A.,
  {Suleimanov} V.,   {Poutanen} J.,  2016, \mn@doi [\aap]
  {10.1051/0004-6361/201628236}, \href
  {https://ui.adsabs.harvard.edu/abs/2016A&A...593A..16T} {593, A16}

\bibitem[\protect\citeauthoryear{{Tsygankov}, {Doroshenko}, {Mushtukov},
  {Lutovinov}  \& {Poutanen}}{{Tsygankov} et~al.}{2018}]{Tsygankov2018}
{Tsygankov} S.~S.,  {Doroshenko} V.,  {Mushtukov} A.~A.,  {Lutovinov} A.~A.,
  {Poutanen} J.,  2018, \mn@doi [\mnras] {10.1093/mnrasl/sly116}, \href
  {https://ui.adsabs.harvard.edu/abs/2018MNRAS.479L.134T} {479, L134}

\bibitem[\protect\citeauthoryear{{Vybornov}, {Doroshenko}, {Staubert}  \&
  {Santangelo}}{{Vybornov} et~al.}{2018}]{Vybornov2018}
{Vybornov} V.,  {Doroshenko} V.,  {Staubert} R.,   {Santangelo} A.,  2018,
  \mn@doi [\aap] {10.1051/0004-6361/201731750}, \href
  {https://ui.adsabs.harvard.edu/abs/2018A&A...610A..88V} {610, A88}

\bibitem[\protect\citeauthoryear{{Wilson-Hodge} et~al.,}{{Wilson-Hodge}
  et~al.}{2018}]{Wilson-Hodge2018}
{Wilson-Hodge} C.~A.,  et~al., 2018, \mn@doi [\apj] {10.3847/1538-4357/aace60},
  \href {https://ui.adsabs.harvard.edu/abs/2018ApJ...863....9W} {863, 9}

\bibitem[\protect\citeauthoryear{{Yoshida} \& {Kitamoto}}{{Yoshida} \&
  {Kitamoto}}{2019}]{Yoshida2019}
{Yoshida} Y.,  {Kitamoto} S.,  2019, \mn@doi [\apj] {10.3847/1538-4357/ab2b3d},
  \href {https://ui.adsabs.harvard.edu/abs/2019ApJ...880..101Y} {880, 101}

\bibitem[\protect\citeauthoryear{{Yoshida}, {Kitamoto}  \& {Hoshino}}{{Yoshida}
  et~al.}{2017}]{Yoshida2017}
{Yoshida} Y.,  {Kitamoto} S.,   {Hoshino} A.,  2017, \mn@doi [\apj]
  {10.3847/1538-4357/aa9023}, \href
  {https://ui.adsabs.harvard.edu/abs/2017ApJ...849..116Y} {849, 116}

\bibitem[\protect\citeauthoryear{{Zheng}, {Liu}  \& {Gou}}{{Zheng}
  et~al.}{2020}]{Zheng2020}
{Zheng} X.,  {Liu} J.,   {Gou} L.,  2020, \mn@doi [\mnras]
  {10.1093/mnras/stz3327}, \href
  {https://ui.adsabs.harvard.edu/abs/2020MNRAS.491.4802Z} {491, 4802}

\makeatother
\end{thebibliography}
